\documentclass[preprint,a4paper,nofootinbib,showkeys]{revtex4-1}

\usepackage{amsmath,amssymb}
\usepackage{enumerate}
\usepackage{amsfonts}

\usepackage{subfigure}
\usepackage[hyperindex]{hyperref}

\usepackage{allpurpose}

\usepackage{tikz}

\usepackage[marginal,perpage,symbol]{footmisc}
\DefineFNsymbols*{dagfirst}{\dag\ddag*{**}{\dag\dag}{\ddag\ddag}}
\setfnsymbol{dagfirst}
\usepackage{slashed} 

\begin{document}

\title{Phantom-like dark energy from quantum gravity}
\date{\today}

\author{Daniele Oriti}
\email[Email: ]{daniele.oriti@physik.lmu.de}
\affiliation{Arnold-Sommerfeld-Center for Theoretical Physics, Ludwig-Maximilians-Universit\"at, Theresienstra\ss e 37, M\"uchen, 80333, Germany}

\author{Xiankai Pang}
\email[Email: ]{Xiankai.Pang@physik.uni-muenchen.de}
\affiliation{Arnold-Sommerfeld-Center for Theoretical Physics, Ludwig-Maximilians-Universit\"at, Theresienstra\ss e 37, M\"uchen, 80333, Germany}

\begin{abstract}
  We analyse the emergent cosmological dynamics corresponding to the mean field hydrodynamics of quantum gravity condensates, in the tensorial group field theory formalism. We focus in particular on the cosmological effects of fundamental interactions, and on the contributions from different quantum geometric modes. The general consequence of such interactions is to produce an accelerated expansion of the universe, which can happen both at early times, after the quantum bounce predicted by the model, and at late times. Our main result is that, while this fails to give a compelling inflationary scenario in the early universe, it produces naturally a phantom-like dark energy dynamics at late times, compatible with cosmological observations. By recasting the emergent cosmological dynamics in terms of an effective equation of state, we show that it can generically cross the phantom divide, purely out of quantum gravity effects without the need of any additional phantom matter. Furthermore, we show that the dynamics avoids any Big Rip singularity, approaching instead a de Sitter universe asymptotically.
\end{abstract}


\maketitle
\tableofcontents
\section{Introduction}
The first problem of any approach to quantum gravity is the identification of candidate fundamental degrees of freedom of spacetime and geometry, and of their quantum dynamics. Candidate quantum gravity formalisms suggest different solutions to this problem, coming from directly quantizing the classical gravitational theory, as in loop quantum gravity \cite{Ashtekar:2021kfp}, in simplicial quantum gravity (e.g.in the dynamical triangulations perspective \cite{Loll:2019rdj}) or in the asymptotic safety scenario \cite{Reuter:2019byg}, from introducing other discrete structures like in causal set theory \cite{Surya:2019ndm}, or from string theory and AdS/CFT perspectives \cite{Blau:1900zza}.

Tensorial group field theories (TGFTs) \cite{Oriti:2011jm,Krajewski:2012aw,Carrozza:2016vsq,Oriti:2014uga,Rivasseau:2012yp, Rivasseau:2016zco, Rivasseau:2016wvy, Delporte:2018iyf} are a generalised quantum field field theory formalism for candidate {\it constituents} of quantum spacetime, that can be represented as elementary simplices, with their fundamental interaction processes represented as simplicial complexes (of one dimension higher). Their perturbative expansion gives in fact a sum over simplicial complexes weighted by model-dependent quantum amplitudes. They are a generalization to higher dimensions of matrix models for 2d quantum gravity. A particularly interesting class of TGFT models, called simply group field theories (GFTs) in the literature, have distinctive quantum geometric aspects. Their fundamental simplices are endowed with group theoretic data encoding their geometric properties and the action is constructed in such a way that the Feynman amplitudes take the explicit form of lattice gravity path integrals, expressed in terms of the same group theoretic variables, on the lattices dual to the TGFT Feynman diagrams, or equivalently spin foam models, when expressed in terms of group representations. The last connection also makes manifest the relation of GFT models with loop quantum gravity \cite{Oriti:2013aqa, Oriti:2014uga}: they can be seen as a second quantized formulation of the kinemetics and dynamics of the fundamental degrees of freedom suggested by canonical loop quantum gravity, i.e. spin networks; spin foam models arise also in loop quantum gravity as the covariant formulation of the dynamics of spin networks.     

The second problem of quantum gravity approaches is to recover the usual description of the universe in terms of a smooth spacetime and fields living on it, and their dynamics governed by (a possibly modified version of) General Relativity and effective quantum field theory. The task is simpler in approaches that in fact {\it start} from the same mathematical structures of effective field theory, like asymptotic safety, or some (even radical) generalization of them, like string theory, which can still make use to a large extent of the usual intuition and tools of spacetime physics (however, they may have then a harder time providing a precise description of the fundamental degrees of freedom underlying spacetime itself). Approaches trying to recover spacetime starting from more abstract, non-spatiotemporal entities find here, instead, a difficult challenge, which is harder the more distant their candidate fundamental entities are from usual spacetime-based fields. The set of such challenges is often referred to as the issue of the {\it emergence of spacetime} in quantum gravity \cite{Oriti:2018dsg}.

Tensorial group field theories belong to this second kind of quantum gravity approaches. In this respect, however, they have the advantage that, despite their fully background independent and non-spatiotemporal character, they can still rely on QFT tools to investigate the emergence of spacetime from their quantum dynamics. This has been one important motivation in the study of RG flows and critical behaviour of a large number of TGFT models \cite{Carrozza:2016vsq,Rivasseau:2012yp, Rivasseau:2016zco, Rivasseau:2016wvy, Finocchiaro:2020fhl, Pithis:2020kio}, having also in mind the way in which matrix models recover 2d continuum Liouville gravity. Group field theories can also rely, for solving the same issue, on the additional quantum geometric data labelling their fundamental quanta and enriching their quantum dynamics. Indeed, while this makes their quantum states and amplitudes more involved, it also provides a guideline for their spatiotemporal interpretation, and makes even the simplest types of approximation schemes geometrically rich enough to be interesting. 

GFT condensate cosmology \cite{Gielen:2013naa,Oriti:2016qtz,Gielen:2016dss, Oriti:2016acw,Pithis:2019tvp,Marchetti:2020qsq,Marchetti:2020umh} is a research programme aiming at the extraction of spacetime physics, in particular cosmology, from group field theories. It is based on the hypothesis that the emergent gravitational physics should be looked for in the hydrodynamic approximation of the full GFT quantum dynamics, and focuses in particular on condensate states, thus treating the universe as a peculiar quantum fluid, made out of the GFT quanta. A large number of recent analyses in this context have shown not only the general viability of this strategy, but also that physically interesting results can be obtained already in the mean field (or Gross-Pitaevskii) approximation. This is also the context of our present work.

Establishing a solid connection between fundamental quantum gravity formalisms and effective spacetime physics is the necessary ingredient to make them testable. This can happen by directly producing new testable predictions about modifications of established theories, like GR or the Standard Model. It can also happen by reproducing some of the existing phenomenological or otherwise simplified models incorporating hypothetical quantum gravity effects or specific features of existing fundamental formalisms (e.g. loop quantum cosmology \cite{Ashtekar:2011ni,Bojowald:2020wuc}, whose dynamics can in fact be reproduced in a specific regime of GFT condensate cosmology). Most current quantum gravity phenomenology is of this type and it is thus waiting for a solid contact with fundamental quantum gravity formalisms. The same is true for existing semi-classical cosmological scenarios for the very early universe: inflationary models, bouncing or emergent universe scenarios. All of them, albeit to a different degree and in very different ways, rely on assumptions about the very early universe that only a more complete theory of quantum gravity can corroborate or replace. 

Quantum gravity effects in cosmology, however, do not need to be confined to the very early universe. In particular, in any emergent spacetime context notions like separation of scales or locality, on which usual effective field theory reasoning is based, are by definition approximate, and one should rather expect that the whole spacetime structure and dynamics, including large scale features, could be discovered to be of direct quantum gravity origin. 

One instance of such large scale cosmological issues that quantum gravity can be expected to resolve is dark energy \cite{Brax:2017idh} (or, closely related to it, the nature of the cosmological constant \cite{Burgess:2013ara}). The nature of the observed cosmic acceleration, and the full characterization of its features, is a main theoretical challenge for modern cosmology. It can be tackled at a more phenomenological level, looking for the semi-classical field-theoretic model that best fits cosmological observations, and indeed there exists a rich zoology of (at least partially) working models who do the job (we include in this category also modified gravity theories).

One such field-theoretic model is so-called phantom dark energy \cite{Caldwell:1999ew}, based on a dynamical new component of the universe (\lq phantom matter\rq) characterized by an equation of state $w < -1$ for a recent part of its history, before tending to the observed $w \approx -1$.
This dark energy evolution, including such phantom crossing, is compatible with current observations and could even be favored by them, e.g. by the recent data on supernova \cite{Shafer:2013pxa,Zhao:2017cud,Wang:2018fng}. 

However, a proper field theoretic modeling of phantom dark energy is challenging, if the phantom field is taken to be a fundamental component of the universe. In fact, the negative kinetic term needed to have $w<-1$ leads to vacuum instability, Lorentz violation or other pathology \cite{Caldwell:1999ew,Carroll:2003st}. Various solutions have been proposed, for example involving several scalar fields \cite{Saitou:2012xw}, but with no conclusive success. Another route to achieve phantom crossing without new fundamental scalar fields is to understand it as a consequence of modified gravity theories, rather than new matter, and this can be accomplished, for example, by suitable $f(R)$ theories \cite{Nojiri:2006be, Bamba:2008hq}.
For the current status of phantom dark energy, see \cite{Ludwick:2017tox}.

An alternative route towards a resolution of the dark energy problem, and in particular for a top-down derivation of 
phantom dark energy, is to obtain it as an effective description of more fundamental quantum gravity dynamics. One can interpret in this way various attempts to model phantom dark energy in string-inspired scenarios that, although still semiclassical, incorporate features of string theory. For example, phantom-like dark energy can be obtained in braneworld models \cite{Chimento:2006ac} and, even more in contact with the fundamental theory, in string gas cosmology \cite{McInnes:2005vp} and in AdS/CFT scenarios \cite{McInnes:2001zw}. Also, the late time acceleration of the universe can also be explained in asymptotically safe cosmology \cite{Anagnostopoulos:2018jdq}, with no need of dark energy or cosmological constants.

In this work we take this route as well, and show that phantom-like dark energy can be obtained naturally in GFT condensate cosmology. It arises as an effective description of the evolution of the universe at late times, in the hydrodynamic approximation of the fundamental quantum dynamics of spacetime constituents, without introducing any kind of special phantom matter, but purely as quantum gravity effect. 

The presentation is organized as follows. We first set up the stage in section \ref{sec:gftcosreview}, by presenting a short review of the GFT formalism \cite{Oriti:2011jm,Krajewski:2012aw} and of GFT condensate cosmology \cite{Oriti:2016acw,Gielen:2016dss,Pithis:2019tvp}. In section \ref{sec:effwsingle}, we introduce the effective equation of state $w$ whose dynamics is the central object of our analysis, summarize the main aspects of phantom dark energy, and recall the results of earlier work concerning the effect of GFT interactions in the emergent cosmological evolution in the single mode case. Then, we move on to our new results. In section \ref{sec:freecondensate}, we consider the early universe dynamics right after the bounce, where the free part in the quantum gravity condensate dominates. We take all quantum geometric modes into account and show that the bounce is followed by a accelerated phase, but this phase is not long lasting in general. The role of GFT interactions is studied in section \ref{sec:lateacc}. We first consider the large volume behaviour of individual modes subject to one interaction term, in section \ref{sec:interactcondensate}. Then, in section \ref{sec:twomodes}, we study how the dynamics of two quantum geometric modes combine to determine the evolution of the universe volume, at late times. We show that the phantom divide can be naturally crossed. We then study the subsequent evolution and how a Big Rip singularity is avoided, leading instead to an asymptotically deSitter universe, in section \ref{sec:bigrip}. Finally, in section \ref{sec:infphantom}, we briefly consider also the case in which each quantum geometric mode is subject to two types of GFT interactions, showing how the late time acceleration phase can have an even richer dynamics, while maintaining the key features of phantom crossing and deSitter asymptotics. Finally, in section \ref{sec:summary} we will summarize our results and give a short outlook toward possible extensions of our work.

\section{GFT condensate cosmology} \label{sec:gftcosreview}
In this section we present some basics of the TGFT formalism and of quantum geometric models (GFTs) for 4d quantum gravity in particular, with a focus on the elements on which the extraction of cosmological dynamics is based. We only include the ingredients that are needed as immediate background of the work presented in this paper. For a more detailed introduction to TGFT we refer to existing reviews \cite{Freidel:2005qe, Oriti:2011jm,Krajewski:2012aw,Oriti:2014uga}. For the basics of GFT cosmology see instead the original work in \cite{Gielen:2013kla,Gielen:2013naa,Oriti:2016qtz,Oriti:2016ueo} and the reviews \cite{Gielen:2016dss,Oriti:2016acw,Pithis:2019tvp}. See also \cite{Marchetti:2020umh,Marchetti:2020qsq} for the use of coherent peaked states for the study of relational observables and their dynamics, and for the discussion of their quantum fluctuations.

\subsection{Group field theory formalism}

GFTs are quantum field theories defined over several copies of a Lie group $G$, which replaces the usual spacetime manifold of standard field theories and does not have, to start with, any spatiotemporal interpretation. In 4d quantum gravity models, the (usually complex) field is a tensorial map $\varphi:~G^{\times 4}\to\mathbb{C},~\varphi(g_v)=\varphi(g_1,\cdots,g_4)$, where the rank of the tensor is chosen equal to the dimension of the spacetime one intends to reconstruct \cite{Oriti:2011jm}. GFTs are understood, in fact, as field theory formulations \emph{of} spacetime, more precisely of the kinematics and dynamics of its fundamental constituents, rather than \emph{on} spacetime as it is the case for usual QFTs. 
The basic quanta of the theory can be depicted as combinatorial 3-simplices, i.e. tetrahedra, labelled by the group-theoretic data, which encode their quantum geometry (assumed to be spacelike). Quantum states and boundary data of such models will correspond to collections of such quanta. 
In the quantum geometric models proposed to date the relevant group manifold is $G=SL(2,\mathbb{C})$  or its rotation subgroup $SU(2)$ (for the Lorentzian signature), since the restrictions that the models impose on the group-theoretic data to ensure a proper geometric interpretation of the simplices allows (in most cases) to map the two formulations of their quantum geometry \cite{Engle:2007wy,Engle:2007uq, Dupuis:2010jn, Finocchiaro:2020xwr}. For details on the quantum geometric conditions, we refer to the cited literature. In the following we will take $G=SU(2)$



Following these geometric restrictions, the field $\varphi(g_v)$ is required to be right invariant, $\varphi(g_vh)=\varphi(g_1h,g_2h,g_3h,g_4h)=\varphi(g_v),~\forall h\in G$, and therefore $\varphi(g_v)\in L^2(G^{\times4}/G)$. A complete and orthonormal basis of $L^2(G^{\times4}/G)$ is given in terms of $SU(2)$ Wigner representation functions contracted by group intertwiners; these are called spin network vertex functions. Such functions are obtained from the Peter-Weyl decomposition of the GFT field
\iea{
  \varphi(g_v)=\sum_{\vec{x}}c_{\vec{x}}\kappa_{\vec{x}},
}
with $c_{\vec{x}}=\int\dd^4 g\varphi(g_v)\kappa_{\vec{x}}(g_v)$ is the projection of field $\varphi(g_v)$ onto basis $\kappa_{\vec{x}}(g_v)$, defined as:
\iea{
  \kappa_{\vec{x}}(g_v)=\sum_{\vec{n}}\left\{\left[\prod_{i=1}^4\sqrt{d(j_i)}D^{j_i}_{m_i n_i}(g_{v_i})\right]\mathcal{I}^{\vec{j},\iota}_{\vec{n}}\right\}\in L^2(G^{\times 4}/G), \label{eq:spinvertexbasis}
}
which are orthonormal under the normalized Haar measure $\int_G\dd g=1$, i.e.,
\iea{
  \int\dd^4g\kappa_{\vec{x}}(g_v)\kappa_{\vec{x}'}(g_v)\equiv\int\prod_i \dd g_{v_i}\kappa_{\vec{x}}(g_v)\kappa_{\vec{x}'}(g_v)=\delta_{\vec{x},\vec{x}'}.
}
These basis functions can be associated graphically to a spin network 4-vertex, i.e. a node with $d=4$ open links associated with $4$ spins $\vec{j}=(j_1,j_2,j_3,j_4)$, together with angular momentum projections $\vec{m}$, and the intertwiner quantum number $\iota$ associated instead to the node itself \cite{Oriti:2013aqa}. Geometrically, one can think the spin network vertex sitting inside the tetrahedron with the $4$ links emanated from the node crossing its $4$ triangular faces. Following the quantization of simplicial geometry for the tetrahedron (whose results are also consonant to the results obtained in the continuum canonical Loop Quantum Gravity context), such spin network states are eigenstates of relevant geometric operators, with the spin labels $j_i$ determining the areas of the four faces, while the intertwiner label specifying the volume of the tetrahedron. 

In the following, we rely on this representation of GFT fields and quantum states.


\paragraph{Second quantization.} GFTs can be dealt with in a second quantization language\footnote{This second quantized formulation, however, is not directly the result of quantizing by standard canonical methods the theory starting from the classical GFT action, due to the lack of external time parameters on which such standard methods would rely. In fact, alternative \lq deparametrized\rq\  formulation of the same GFTs (after additional \lq matter\rq\  degrees of freedom have been included; see the following sections) exist \cite{Wilson-Ewing:2018mrp, Gielen:2019kae}. This timeless formalism can also be derived by more standard canonical quantization methods from a \lq frozen\rq\  perspective \cite{Gielen:2021dlk}, looking at the GFT model as a peculiar constrained system.} by promoting the fields and their modes $c_{\vec{x}}$ to operators \cite{Oriti:2013aqa}, 
\iea{
  \hat{\varphi}(g_v)=\sum_{\vec{x}}\hat{c}_{\vec{x}}\kappa_{\vec{x}}(g_v),~\hat{\varphi}^\dag(g_v)=\sum_{\vec{x}}\hat{c}_{\vec{x}}^\dag\bar{\kappa}_{\vec{x}}(g_v), 
}
where the annihilation and creation operator $\hat{c}_{\vec{x}}$ and $\hat{c}_{\vec{x}}^{\dag}$, create/annihilate spin network nodes (or, equivalently, tetrahedra) labeled by $\vec{x}=(\vec{j},\vec{m},\iota)$, and satisfy the commutation relations
\iea{
  \left[\hat{c}_{\vec{x}},\hat{c}^\dag_{\vec{x}'}\right]=\delta_{\vec{x},\vec{x}'},~\left[\hat{c}_{\vec{x}},\hat{c}_{\vec{x}'}\right]=\left[\hat{c}^\dag_{\vec{x}},\hat{c}^\dag_{\vec{x}'}\right]=0.
}

The vacuum $\ket{0}$, which is the state with no spacetime structure (geometrical or topological), is defined by  $\hat{c}_{\vec{x}}\ket{0}=0, ~\forall \vec{x}$. By acting the creation operator $\hat{c}^\dag_{\vec{x}}$ repeatedly on $\ket{0}$ we can construct the many-body states as usual, leading to the Fock space
\ieas{
  \mathcal{F}(\mathcal{H})=\operatorname*{\oplus}_{N=0}^\infty sym\left\{\mathcal{H}^{(1)}\otimes\cdots\otimes\mathcal{H}^{(N)}\right\},~\mathcal{H}=L^2\left(G^{\times 4}/G\right),
}
where $N$ denotes the number of tetrahedra in each sector of the Fock space. Here, bosonic statistics is assumed. Extended topological structures, corresponding to simplicial complexes formed by glued tetrahedra, or equivalently by graphs formed by connected spin network vertices, can be put in precise correspondence with entangled many-body states inside the Fock space, with the graph structure encoding exactly the entanglement pattern among fundamental degrees of freedom \cite{Colafranceschi:2020ern}.

For this formalism to be a compelling formulation of quantum gravity, our universe, including its dynamical spacetime geometry, should be shown to emerge from the quantum dynamics of these more abstract fundamental entities. This is the focus of the present work.

Second quantized versions of the various quantum geometric operators can be then constructed.

In the following we need the volume operator, which is diagonal in the spin network basis with matrix elements depending on the intertwiner label $\iota$. Therefore we can write \cite{Marchetti:2020umh}
\iea{
  \hat{V}=\sum_{\vec{x},\vec{x}'}V(\iota,\iota')\delta_{\vec{x}-\{\iota\},\vec{x}'-\{\iota'\}}\hat{c}_{\vec{x}}^\dag\hat{c}_{\vec{x}'}.
}

\paragraph{Coupling to a scalar field.} In a diffeomorphism invariant context, a convenient strategy to define time evolution is a relational one in which appropriate internal dynamical degrees of freedom of the theory are used as a clock, with respect to which the evolution of the others is defined \cite{Tambornino:2011vg,Hoehn:2019owq}. In many applications, and in GFT cosmology in particular, the role of a clock is played by a massless non-interacting (and minimally coupled) scalar field. Such scalar field degrees of freedom is added to the quantum geometric ones in the fundamental definition of the GFT model. The first step is to extend the definition of GFT field to be the map $\varphi: G^{\times 4}\times \mathcal{R}\to \mathcal{C}$, and then the GFT action should be extended to include appropriate coupling of the new degrees of freedom. The main guideline for constructing such extended dynamics is in fact the same as for the pure geometry models: the GFT model is defined in such a way that its perturbative expansion produces a sum over simplicial complexes weighted by an appropriate discrete path integral for gravity, now coupled to a massless non-interacting scalar field \cite{Gielen:2013naa,Oriti:2016qtz, Li:2017uao}. 
Let us stress that, while the interpretation of the new degrees of freedom, just like that of the quantum geometric ones, is guided by the role they play at the discrete level corresponding to GFT quanta and Feynman amplitudes, their actual physical meaning and properties should be determined by the role they play at some effective continuum level. The GFT cosmology programme is exactly aimed at extracting such effective description and controlling the emerging physics of these quantum gravity models.

After quantization, the field operators will also be $\phi$ dependent, in particular the commutation relation between annihilation and creation operator becomes
\iea{
  \left[\hat{c}_{\vec{x}}(\phi),\hat{c}^\dag_{\vec{x}'}(\phi')\right]=\delta_{\vec{x},\vec{x}'}\delta(\phi'-\phi),~\left[\hat{c}_{\vec{x}}(\phi),\hat{c}_{\vec{x}'}(\phi')\right]=\left[\hat{c}^\dag_{\vec{x}}(\phi),\hat{c}^\dag_{\vec{x}'}(\phi')\right]=0.
}

Correspondingly, the definition of other observables will include their dependence on scalar field degrees of freedom. For example, the volume operator, counting the contribution from each GFT quantum, becomes: 
\iea{
  \hat{V}=\int\dd\phi\hat{V}(\phi)=\int\dd\phi\sum_{\vec{x},\vec{x}'}V(\iota,\iota')\delta_{\vec{x}-\{\iota\},\vec{x}'-\{\iota'\}}\hat{c}_{\vec{x}}^\dag(\phi)\hat{c}_{\vec{x}'}(\phi). \label{eq:deftotalvolume}
}

The relational strategy would then suggest to look for a definition of a relational observable corresponding to the volume of the universe {\it at given clock time}, with the role of clock played by the scalar field, and a first definition could be given by the quantity $\hat{V}(\phi)$ entering the above expression. Indeed, this has been the definition adopted in much of the GFT cosmology literature. Recently, an effective relational strategy has been proposed, in which relational observables correspond to the expectation values of the generic GFT operators in appropriately selected \lq clock-peaked\rq\   states. We are going to illustrate this effective strategy in the following, after discussing the dynamical aspects of the theory.


\paragraph{Dynamics.} Classically, the dynamics of a given GFT model is specified by the action
\iea{
  S(\bar{\varphi},\varphi)&=& \int\dd g_{v_1}\dd g_{v_2}\bar{\varphi}(g_{v_1})\varphi(g_{v_2})K(g_{v_1},g_{v_2})\nonumber \\
  &&-\sum_{n,m}^\infty\lambda_{n+m}\int\left[(\dd g_v)^{m}(\dd h_v)^{n}\prod_{i=1}^m\bar{\varphi}(g_{v_i})\prod_{j=1}^n\varphi(h_{v_j})V_{n+m}(g_{v},h_{v})\right],
}
where $K(g_{v_1},g_{v_2})$ and $V_{n+m}(g_v,h_v)=V_{n+m}(g_{v_1},\cdots,g_{v_m},h_{v_1},\cdots,h_{v_n})$ are kinetic and interaction kernels respectively. We have adopted a notation reminiscent of quantum many-body physics, indicating that different interactions involving varying numbers of \lq spacetime atoms\rq\  are possible, and restricted to the case of pure quantum geometric data for simplicity of notation. The interaction kernels are generically non-local with respect to such quantum geometric data, in the sense that field arguments are not simply identified at the interaction. When scalar field degrees of freedom are present, on the other hand, typical interaction kernels are going to be local in them. 
The quantum dynamics can be extracted from the partition function\footnote{This partition function can be seen as the result of rewriting in path integral form a \lq generally covariant equilibrium partition function\rq\  of quantum statistical type for a system of quantized simplices; see \cite{Chirco2018b,Kotecha:2020oxz}.} 
\ieas{
  Z=\int\mathcal{D}\varphi\mathcal{D}\bar{\varphi}\ee^{-S(\bar{\varphi},\varphi)},
}
from which we get the Schwinger-Dyson equations \cite{Gielen:2013naa,Gielen:2016dss}
\iea{
  0=\int\mathcal{D}\varphi\mathcal{D}\bar{\varphi}\frac{\delta}{\delta\bar{\varphi}}\left(O(\bar{\varphi},\varphi)\ee^{-S(\bar{\varphi},\varphi}\right)=\left\langle\frac{\delta O(\bar{\varphi},\varphi)}{\delta \bar{\varphi}}-O(\bar{\varphi},\varphi)\frac{\delta S(\bar{\varphi},\varphi)}{\delta \bar{\varphi}}\right\rangle,
}
where the vacuum expectation value $\braket{\cdots}$ is defined as
\ieas{
  \langle O(\bar{\varphi},\varphi)\rangle=\int\mathcal{D}\varphi\mathcal{D}\bar{\varphi}O(\bar{\varphi},\varphi)\ee^{-S(\bar{\varphi},\varphi)}.
}

When quantum fluctuations are small, a mean field approximation is expected to be valid. This means that at the leading order we need only consider the simplest one in the series of Schwinger-Dyson equations
\ieas{
  \left\langle\sigma\left|\frac{\delta \hat{S}(\hat{\varphi}^\dag,\hat{\varphi})}{\delta\hat{\varphi}^\dag}\right|\sigma\right\rangle=0
}
for any state $\ket{\sigma}$. In particular, if $\ket{\sigma}$ is eigenstate of field operator, i.e., $\hat{\varphi}(g_v,\phi)\ket{\sigma}=\sigma(g_v,\phi)\ket{\sigma}$ for some eigenfunction $\sigma(g_v,\phi)$, the dynamics can be expressed as equation of motion for $\sigma(g_v,\phi)$ obtained from the effective action
\iea{
  S(\bar{\sigma},\sigma)=\langle\sigma|S(\hat{\varphi}^\dag,\hat{\varphi})|\sigma\rangle.
}

This approximation can also be seen as corresponding to approximating the full quantum effective action of the field theory with its classical one, since the resulting equations of motion are the ones obtained from the classical action replacing the GFT field with the function $\sigma(g_v,\phi)$. In the context of quantum many-body system, specifically quantum liquids, this is the Gross-Pitaevskii approximation of the condensate hydrodynamics for the condensate wavefunction $\sigma(g_v,\phi)$.  

Now we discuss how this approximation plays out for a special class of condensate wavefunctions, leading to an effective definition of relational observables and to an emergent cosmological dynamics. 

\subsection{GFT condensate cosmology}

\paragraph{Coherent peaked states.} In our framework, the evolution of the universe can expressed as the change of a spatial slice of spacetime with respect to relational time $\phi$. In order to  introduce this dependence of observables on the value of our clock, we work with states peaked on a fixed relational time $\phi_0$ \cite{Marchetti:2020umh}. The same states should support the contribution to such observables of large number of fundamental GFT quanta, which is expected to necessary for a good continuum approximation \cite{Oriti:2016qtz}. These two considerations lead to the use of the coherent peaked states (CPS): 
\iea{
  \ket{\sigma_\varepsilon;\phi_0,\pi_0}=\mathcal{N}(\sigma)\exp\left(\int(\dd g)^4\dd\phi \sigma_\varepsilon(g_v,\phi;\phi_0,\pi_0)\hat{\varphi}^{\dag}(g_v,\phi)\right)\ket{0}, \label{eq:annidef}
}
with $\mathcal{N}(\sigma)$ is some normalization constant and $\ket{0}$ is the vacuum state. The condensate wavefunction $\sigma_\varepsilon(g_v,\phi;\phi_0,\pi_0)$ is peaked on $\phi=\phi_0$ and can be written as \cite{Marchetti:2020qsq}
\iea{
  \sigma_\varepsilon(g_v,\phi;\phi_0,\pi_0)=\eta_\varepsilon(\phi-\phi_0,\pi_0)\tilde{\sigma}(g_v,\phi),
}
where $\eta_\varepsilon(\phi-\phi_0,\pi_0)$ is a \emph{peaking function} (usually taken as a Gaussian, see equation (52) in \cite{Marchetti:2020umh}) around $\phi_0$ with a typical width given by $\varepsilon$, and $\pi_0$ is a further parameter controlling the fluctuations of the operator corresponding to the conjugate momentum of the scalar field $\phi$.  The \emph{reduced condensate function} $\tilde{\sigma}(g_v,\phi)$, which is the actual dynamical variable in the hydrodynamic approximation, does not modify the peaking property of $\sigma_\varepsilon(g_v,\phi;\phi_0,\pi_0)$, determined by $\eta_\varepsilon(\phi-\phi_0,\pi_0)$. It remains true, of course, that the condensate state \ref{eq:annidef} is an eigenstate of GFT field operator
\iea{
  \hat{\varphi}(g_v,\phi)\ket{\sigma_\varepsilon;\phi_0,\pi_0}=\sigma_\varepsilon(g_v,\phi;\phi_0,\pi_0)\ket{\sigma_\varepsilon;\phi_0,\pi_0}.
}
One further condition imposed on the condensate wavefunction, motivated by geometric considerations \cite{Marchetti:2020umh, Gielen:2013naa, Oriti:2016qtz,Gielen:2014ila}, is invariance under both right and left diagonal group actions
\iea{
  \tilde{\sigma}(h g_v k,\phi)=\tilde{\sigma}(g_v,\phi),~\forall h,~k\in SU(2).
}

\paragraph{Imposing isotropy.}
We are interested in reproducing the cosmological dynamics of homogeneous and isotropic universes from the GFT condensate hydrodynamics, fully encoded in the evolution of the universe volume. Therefore, we impose one further restriction on the condensate wavefunction, i.e. isotropy.
This becomes the requirement that the wave function $\sigma(g_I,\phi)$ only has support over equilateral tetrahedra, corresponding to the restriction of all spin labels, i.e. the areas of its boundary triangles, being equal and the volume eigenvalue being the maximal one allowed by this choice of triangle areas. Taking into account also the left and right invariance, the condensate wavefunction is then of the form \cite{Oriti:2016qtz}
\iea{
  \tilde{\sigma}(g_I,\phi)=\sum_j\tilde{\sigma}_j(\phi)\bar{\mathcal{I}}^{j,\iota_{+}}_{\vec{m}}\mathcal{I}^{j,\iota_{+}}_{\vec{n}}d(j)^2\prod_{l=1}^4D^j_{m_ln_l}(g_l), \label{eq:sigmadecomp}
}
where we write $j$ for $\vec{j}=(j_1,j_2,j_3,j_4)=(j,j,j,j)$, and similarly for $\vec{m},~\vec{n}$; $\mathcal{I}^{j,\iota_+}_{\vec{m}}$ is the intertwiner labeled  by $\iota$, $d(j)=2j+1$ is the dimension of the spin $j$ representation and $D^{j}_{m_ln_l}(g_l)$ are the Wigner representation functions. The dependence on relational time is then only encoded in $\sigma_j(\phi)$ for each mode. Note that, given the definition \eqref{eq:annidef} of annihilation operator $\hat{c}_{\vec{x}}$, we have 
\iea{
  \hat{c}_{\vec{x}}(\phi)\ket{\sigma_\varepsilon;\phi_0,\pi_0}=\eta_\varepsilon(\phi-\phi_0,\pi_0)\tilde{\sigma}_j(\phi)\bar{\mathcal{I}}^{j,\iota_+}_{\vec{m}}\ket{\sigma},
}
i.e., only for $j_1=j_2=j_3=j_4=j$ the action of $\hat{c}_{\vec{x}}$ with $\vec{x}=(\vec{j},\vec{m},\iota)$ is not vanishing. 

\paragraph{Effective dynamics.}

Having fixed the peaking function $\eta_\varepsilon(\phi-\phi_0,\pi_0)$, the dynamics of the condensate is encoded in the evolution of the reduced condensate function $\tilde{\sigma}(g_v,\phi)$. Furthermore, at mean field level the dynamics can be extracted from an effective action. This reads \cite{Marchetti:2020umh}:
\iea{
  S(\bar{\tilde{\sigma}},\tilde{\sigma})&=&\int\dd\phi_0\langle\sigma_\varepsilon;\phi_0,\pi_0|S(\hat{\varphi}^\dag,\hat{\varphi})|\sigma_\varepsilon;\phi_0,\pi_0\rangle, \nonumber \\
  &=&\int\dd\phi_0\left\{\sum_j\left[\bar{\tilde{\sigma}}_j(\phi_0)\tilde{\sigma}''(\phi_0)-2\ii \tilde{\pi}_0\bar{\tilde{\sigma}}_j(\phi_0)\tilde{\sigma}_j'(\phi_0)-\xi_j^2\bar{\tilde{\sigma}}_j(\phi_0)\tilde{\sigma}_j(\phi_0)\right]+\mathcal{V}(\bar{\tilde{\sigma}},\tilde{\sigma})\right\}, \label{eq:condaction}
}
where $\displaystyle \tilde{\pi}_0=\frac{\pi_0}{\varepsilon \pi_0^2-1}$, $\xi_j$ is an effective parameter encoding the details of the kinetic term of the fundamental GFT action (in the isotropic restriction), and the derivatives $'$ denote derivatives with respect to $\phi_0$. Finally,  $\mathcal{V}(\bar{\tilde{\sigma}},\tilde{\sigma})$ is the interaction kernel, also determined by the underlying GFT model. 
We refer to \cite{Marchetti:2020umh}, and references cited therein, for more details.

The interaction term for quantum geometric GFT models remains quite involved also in the isotropic restriction, and the corresponding dynamics is difficult to handle even at this mean field level. For this practical reason, most analyses so far have neglected the contribution coming from such interaction terms, which are expected to be anyway subdominant with respect to the kinetic part\footnote{This is also needed, in fact, for the perturbative form of the GFT quantum dynamics, where the connection with spin foam models and lattice gravity path integral is established, to be of any relevance.}. In this work, on the other hand, we want to focus exactly on how these interaction terms affect the effective cosmological dynamics, especially at late times. 

For doing so, we adopt a rather phenomenological approach, modelling these interactions with a simple, rather general form, used also in previous work \cite{deCesare:2016rsf}:
\iea{
  \mathcal{V}(\bar{\tilde{\sigma}},\tilde{\sigma})=\sum_j\left(\frac{2\lambda_j}{n_j}|\tilde{\sigma}_j(\phi_0)|^{n_j}+\frac{2\mu_j}{n_j'}|\tilde{\sigma}_j(\phi_0)|^{n_j'}\right),\label{eq:intkernel}
}
where $\lambda_j$ and $\mu_j$ are interaction couplings correspond to each mode $j$ satisfy that $|\mu_j|\ll|\lambda_j|\ll|m_j^2|$ and we assume that $n_j'>n_j>2$. 
Albeit definitely simpler than full-blown quantum geometric models, this choice still captures several relevant features of the same, and hopefully key aspects of what we may expect to be universal effective behaviour. We emphasis that at this stage our effective action is not derived from some underlying GFT model. We choose the interaction kernel to be equation \eqref{eq:intkernel} as it is easy to handle and also has a similar structure of some microscopic GFT theories, such as the one corresponding to EPRL model \cite{Oriti:2016qtz}. In this sense, any GFT model that can reproduce such effective action (under mean-field approximation or with some quantum corrections) would lead to the same evolution of the universe that we will explore below.

Note that there is no cross term among different modes in the action, therefore the equations of motion for different $j$ decouple. This would simplify the analysis quite a bit. At the same time, some models like the EPRL model decouple different modes in the isotropic restriction \cite{Oriti:2016qtz}. For more general GFT actions, different modes can couple to each other (for example \cite{Baratin:2011hp}, in the Riemannian setting) and the analysis would be more involved, we leave the study of behaviour of coupled modes for future work. 

In \cite{deCesare:2016rsf} this kind of interactions has been studied in the case in which only a single spin mode contributes, and it has been shown that they affect the effective universe dynamics in interesting ways. For example it allows to obtain an inflationary phase in the early universe. 

We will improve on this earlier work by considering the contribution of more than one mode, and show that the result is even more interesting; in particular, we will show that we can obtain an effective dark energy dynamics at late times, produced directly from the underlying quantum gravity dynamics, without introducing any kind of additional matter-like field. 

Before analyzing the resulting dynamics for the universe volume, obtained from this effective condensate action, let us recast it in a more convenient hydrodynamic form\footnote{Since the equation of motion only depend on $\phi_0$ and thus there is no risk of confusion, for notation simplicity in the following we will drop the subscript $0$ and use $\phi$ to represent the relational time.}. 

Varying the action \eqref{eq:condaction} with respect to $\bar{\tilde{\sigma}}_j$ we get \cite{Marchetti:2020umh,deCesare:2016axk}
\iea{
  \tilde{\sigma}_j''-2\ii \tilde{\pi}_0\tilde{\sigma}_j-\xi_j^2\tilde{\sigma}_j+2\lambda_j|\tilde{\sigma}_j|^{n_j-2}\tilde{\sigma}_j+2\mu_j|\tilde{\sigma}_j|^{n_j'-2}\tilde{\sigma}_j=0.
}
 Decomposing $\tilde{\sigma}_j(\phi)=\rho_j(\phi)\exp[\ii \theta_j(\phi)]$ with real $\rho_j$ (condensate density) and $\theta_j$ (condensate phase), then last equation gives two equations for real and imaginary parts respectively. Using a global $U(1)$ symmetry of our equation (and effective action), the imaginary part can be expressed as a total derivative, $Q_j'=0$, with \cite{Oriti:2016qtz,Marchetti:2020umh}
\iea{
  Q_j=(\theta_j'-\tilde{\pi}_0)\rho_j^2
} being the corresponding conserved quantity.
The remaining equation becomes \cite{deCesare:2016rsf,Marchetti:2020umh}
\iea{
  \rho_j''-\frac{Q_j^2}{\rho_j^3}-m_j^2\rho_j+\lambda_j\rho_j^{n_j-1}+\mu_j\rho_j^{n_j'-1}=0, \label{eq:rhojpp}
}
where $\displaystyle m_j^2=\xi_j^2-\tilde{\pi}_0^2$ is now both a function of the fundamental parameters of the model (through $\xi_j$) and of the parameter $\epsilon$ characterizing the non-ideal nature of our clock. This equation can be directly integrated once, which gives another conserved quantity \cite{Oriti:2016qtz}, as a result of the symmetry of the system under \lq\lq clock-time translation\rq\rq\  \cite{deCesare:2016rsf,Marchetti:2020umh},
\iea{
  E_j=\frac{1}{2}(\rho_j')^2-\frac{1}{2}m_j^2\rho_j^2+\frac{Q_j^2}{2\rho_j^2}+\frac{\lambda_j}{n_j}\rho_j^{n_j}+\frac{\mu_j}{n_j'}\rho_j^{n_j'}. \label{eq:defej}
}
From this equation for the condensate density we will now derive an effective evolution equation for the volume of the universe in relational time.

\subsection{Volume dynamics} \label{sec:volumedynamics}

 The expectation value of the volume operator $\hat{V}$ in the condensate state $\ket{\sigma}$ takes the form  \cite{Oriti:2016qtz, Marchetti:2020qsq}
\iea{
  V(\phi_0)&=&\braket{\sigma_\varepsilon;\phi_0,\pi_0|\hat{V}|\sigma_\varepsilon;\phi_0,\pi_0} \nonumber \\
  &=&\braket{\sigma_\varepsilon;\phi_0,\pi_0|\int\dd\phi\sum_{\vec{x},\vec{x}'}V(\iota,\iota')\delta_{\vec{x}-\{\iota\},\vec{x}'-\{\iota'\}}\hat{c}_{\vec{x}}^\dag(\phi)\hat{c}_{\vec{x}'}(\phi)|\sigma_\varepsilon;\phi_0,\pi_0} \nonumber \\
  &\approx&\sum_jV_j\rho_j(\phi_0)^2, \label{eq:totalVcondensate}
} where 
$\rho_j=|\sigma_j|$ is the modulus of reduced condensate function $\tilde{\sigma}$, $V_j\propto l_p^3j^{3/2}$ is the volume contribution from each quantum (tetrahedron) in the spin $j$ representation,and we have used the intertwiner normalization condition $\sum_{\vec{m}}\mathcal{I}^{j,\iota_+}_{\vec{m}}\bar{\mathcal{I}}^{j,\iota'_+}_{\vec{m}}=\delta_{\iota,\iota'}$. The approximation amounts to keeping only the dominant contribution to the saddle point approximation of the peaking function coming from our choice of state (we restored the subscript $0$ for given relational time $\phi_0$ for the moment to avoid confusion) \cite{Marchetti:2020umh}. 

The dynamics of the universe volume can now be obtained by differentiating $V(\phi)$ respect to relational time and then substituting the equations \eqref{eq:rhojpp} and \eqref{eq:defej} for $\rho_j$, writing them in the form of modified FLRW equations \cite{Oriti:2016qtz}
\iea{
  \left(\frac{V'}{3V}\right)^2&=&\left[\frac{2\sum_j V_j\sqrt{2E_j\rho_j^2-Q_j^2+m_j^2\rho_j^4-\frac{2}{n_j}\lambda_j\rho_j^{n_j+2}-\frac{2}{n'_j}\mu_j\rho_j^{n'_j+2}}}{3\sum_k V_k\rho_k^2}\right]^2, \label{eq:vpsquare}\\
  \frac{V''}{V}&=&\frac{2\sum_j V_j\left[2E_j+2m_j^2\rho_j^2-\left(1+\frac{2}{n_j}\right)\lambda_j\rho_j^{n_j}-\left(1+\frac{2}{n_j'}\right)\mu_j\rho_j^{n_j'}\right]}{\sum_k V_k\rho_k^2}. \label{eq:vpp}
}
Note that we only consider the expansion phase, so we chose the sector $\rho_j'\geq0$ when we substituted equation \eqref{eq:defej}. 

We will focus on these two equations in the following discussion, by writing them in the form of standard cosmological equations in terms of an effective equation of state in relational language, and analyzing its behaviour when the universe volume grows. The late time behaviour of the model, we will see, is particularly interesting and can naturally describe a dark energy-driven acceleration, of pure quantum gravity origin. Before doing so, we mention a couple of key features of the dynamics, studied first in \cite{Oriti:2016qtz} and \cite{deCesare:2016rsf}.

\paragraph{Bounce.} At very early time, the volume is small (in fact, so is the modulus $\rho_j$ for each mode, and the dynamics can be well approximated by the free evolution, ignoring the contribution from interactions. One can verify that, as long as one of the $Q_j$'s is non-zero, the corresponding $\rho_j$ cannot reach $0$, so that the square root in equation \eqref{eq:vpsquare} is real. Consequently, the total volume will not reach $0$ and the classical big bang singularity is replaced by a bounce \cite{Oriti:2016qtz}. In fact, even if all the $Q_j$'s vanish, the bouncing scenario is obtained for a large class of parameters (those for which \ref{eq:defej} does not vanish for at least one $j$), and can thus be considered a rather general, albeit not universal, consequence of the quantum gravity dynamics described by the GFT model \cite{Marchetti:2020umh}. 
\paragraph{Classical limit.} As the volume grows, but before the GFT interactions become relevant, we reach a regime where the dynamics can be well approximated by the FLRW equation in the presence of a free massless field \cite{Oriti:2016qtz,Marchetti:2020umh}. In fact, when $\rho_j$ is large $\rho_j^2\gg E_j/m_j^2$ and $\rho_j^3\gg Q_j^2/m_j^2$ while not so large such that $|\mu_j|\rho_j^{n_j'-2}\ll|\lambda_j|\rho_j^{n_j-2}\ll m_j^2$, equations \eqref{eq:vpsquare} and \eqref{eq:vpp} can be approximated by
\ieas{
 \left(\frac{V'}{3V}\right)^2&=&\left(\frac{2\sum_j V_jm_j\rho_j^2}{3\sum_k V_k\rho_k^2}\right)^2, ~
  \frac{V''}{V}=\frac{\sum_j V_j\left(4m_j^2\rho_j^2\right)}{\sum_k V_k\rho_k^2}. 
}
If at least for a dominant spin mode $m_{\tilde{j}} \approx const$\footnote{Note that this is just a sufficient condition, not a necessary one.}, we can {\it define} $m_{\tilde{j}}^2\equiv 3\pi G$ in terms of an effective dimensionless Newton constant $G$,  and the equation takes the form of the FLRW equation with a free massless scalar field in relational time \cite{Oriti:2016qtz,Marchetti:2020umh}
\ieas{
  \left(\frac{V'}{V}\right)^2=\frac{V''}{V}=12\pi G.
}
Furthermore, it can be shown that in the free case the lowest spin mode $j_0$ will dominate quickly \cite{Gielen:2016uft}, therefore it is sufficient that $m_{j_0}^2=3\pi G$ to recover the FLRW equation. 
Thus we see that one can also obtain the correct classical limit at large volumes from the effective GFT condensate hydrodynamics, at least as long as the GFT interactions remain subdominant.

Let us also stress that the above results have been obtained by several different strategies, beyond the specific one we illustrated above, thus confirming their solidity \cite{Wilson-Ewing:2018mrp, Gielen:2019kae}. Moreover, quantum fluctuations of the relevant geometric observables can be analysed in some detail \cite{Marchetti:2020qsq}; the analysis confirms that fluctuations are naturally suppressed at late times, thus the semiclassical limit is reliable, and it allows to put precise constraints on the range of values of the various parameters in the model, for which the same quantum fluctuations remain under control in the bounce region at early times, and for which the relational evolution remains valid as well, i.e. the chosen clock remains a good one. 

\

The important issue becomes, then, how the GFT interactions modify the effective dynamics. This is the issue we tackle in this work, extending the first analyses of this issue, performed in \cite{Pithis:2016cxg,Pithis:2016wzf,deCesare:2016rsf}.

\section{Effective equation of state} \label{sec:effwsingle}
A convenient way to capture the relevant features of the effective cosmological dynamics, that we can extract from the GFT condensate hydrodynamics, is to express it in terms of an effective matter component, in turn described entirely by its equation of state.

In a homogeneous universe, for example, the matter content is assumed to be a perfect fluid and can be characterized by its energy density $\rho$ and pressure $p$ in a comoving frame. The fluid then couples to the geometry, determining the cosmological evolution, through its equation of state $w=p/\rho$. For example, if the expanding universe is dominated by a fluid with $w<-1/3$, then the expansion will be accelerating. Current cosmological observations give a value $w\simeq-1$, thus indeed an accelerating expansion of the observable universe, while the usual matter content from the standard model would give $w=1/3$ for relativistic particles and $w=0$ for non-relativistic particles. Moreover, while a small positive cosmological constant could reproduce this value, one would be left to explain how the value of the cosmological constant is chosen, how it is affected by the quantum dynamics of matter and its interaction with (quantum) gravity, and, more important, how this value changes over time, since a simple constant value is not obviously compatible with what we know about cosmological evolution. This is, in summary, the problem of dark energy \cite{Martin:2012bt}. 

We now express our emergent cosmological dynamics in the same language, appropriately recast in terms of relational clock evolution. 

For a homogeneous and isotropic metric with scale factor $a(t)$, the Hubble parameter can be given by $H=\dot{a}/a$ with the $\dot{~}$ represents the derivative respect to comoving time $t$. Then the effective equation of state can be defined as $w=-1-2\dot{H}/(3H^2)$. In the GFT (and more generally, quantum gravity) context, we cannot rely at the fundamental level on any time coordinate or direction. We can use, instead, a relational definition of time in terms of a physical clock, for example a free massless scalar field $\phi$, as discussed in section \ref{sec:gftcosreview}. In appendix \ref{sec:effw} we show that using this definition of relational time, the equation of state $w$ can be defined by
\iea{
  w=3-\frac{2VV''}{(V')^2}, \label{eq:effwdef}
}
where $V$ is the total volume and the $'$ indicates the derivative with respect to the relational time $\phi$, and we chose the time gauge, in which the volume $V=a^3$ for scale factor $a$.

Using this effective equation of state, all the effects produced on the evolution of the universe by the underlying quantum gravity dynamics can be described as if they were due to some effective matter field $\psi$ satisfying $w_\psi\equiv p_\psi/\rho_\psi=w$, with $p_\psi$ and $\rho_\psi$ its pressure and energy density, respectively. 

We emphasize again that the field $\psi$ introduced this way is just a convenient rewriting of what remains due to the fundamental quantum gravity dynamics. As such, it is not required to possess the usual features of well-behaved matter field theories defined on cosmological backgrounds, nor the desiderata of effective field theory. For the same reason, we will not discuss possible Lagrangians for $\psi$, or dwell any further into its properties {\it qua matter field}. 

One main advantage of introducing the fictitious field $\psi$, beside making the analysis of the volume evolution more practical, is that it helps to gain an intuitive understanding of quantum effects on geometry, or more precisely, on the scalar curvature, which is a rather tricky observable to define and compute in the fundamental quantum geometric GFT context. In fact, suppose the energy-momentum tensor of field $\psi$ is given by $T_{\mu\nu}$, then tracing the Einstein equation we see that the scalar curvature in a universe dominated by $\psi$ can be given by $R=-T^\mu_\mu=-(1+3w_\psi)\rho_\psi$, where we used the fact that $T_\mu^\mu=\rho_\psi+3p_\psi$ in the comoving frame. In particular, this helps identifying potentially singular regimes. For example, if $\rho_\psi\to\infty$, we see that the scalar curvature diverges as well (except for $w\neq-1/3$, which, as we can see in subsection \ref{sec:evopsi}, will not lead to a divergent energy density anyway); this is the so-called \emph{Big Rip} singularity, which is relevant for dark energy models \cite{Caldwell:2003vq,Nojiri:2003vn,Nojiri:2005sx}, and on which we are going to have more to say in the following. 

\subsection{The evolution of $\psi$} \label{sec:evopsi}

Now we recall the evolution of an effective field $\psi$ endowed with the equation of state $w$. We stress once more that we intend this to be only an illustration of which properties a field of this type would have in the context of standard General Relativity and effective (quantum) field theory, making use of all the auxiliary structures (topological manifold, coordinates, gauge conditions, etc) that are useful tools in such context. It is not a determination of the physical properties of a physical field, corresponding to fundamental degrees of freedom and observables of our quantum gravity formalism, but only an effective rewriting of quantum \lq pregeometric\rq\   gravity degrees of freedom, which are not described in terms of similar auxiliary structures. For example, we could {\it define} an energy density for the effective field $\psi$ from the equation of state $w$ and the universe volume $V$ and study its properties, but there is no independent fundamental observable corresponding to it, in the GFT algebra of (2nd quantized) observables.

Having clarified this important point, the energy density $\rho_\psi$ satisfies the conservation equation $\dot{\rho}_\psi+3H(1+w)\rho_\psi=0$. Using the standard definition of Hubble parameter in time gauge $H=\dot{a}/a=\dot{V}/(3V)$, this equation can be rewritten as
\iea{
  \frac{\dd \rho_\psi}{\dd V}+\frac{1+w}{V}\rho_\psi=0\qquad , \label{eq:psiconservation}
}
which can indeed be taken as a definition of the energy density in terms of quantities corresponding to GFT observables. 
For constant $w$, equation \eqref{eq:psiconservation} can be easily solved and the solution is given by
\ieas{
  \rho_\psi=\frac{\rho_{\psi0}}{V^{1+w}},
}
with the $\rho_{\psi0}$ is the constant of integration. For $w>-1$, the energy density $\rho_\psi$ decreases as the volume grows, and tends to vanish  when volume is large, i.e., we expect, at late times; for $w=-1$, the energy density is a constant, corresponding to a cosmological constant, and would tend to dominate over any other fluids with $w>-1$ at late times; for $w<-1$, on the other hand, $\rho_\psi$ increases as the volume becomes larger, and would tend to diverge for $V\to\infty$. Using the Einstein's equations (but the conclusion would hold with most generalizations of GR), we would then find that the scalar curvature would diverge as well, i.e. $R=-(1+3w_\psi)\rho_\psi \to\infty$. This is referred to as a Big Rip singularity. 

The above discussion gives a first intuition for the possible late time evolution of our universe, and of various issues constituting the dark energy problem. It should be clear, however, that things are so simple only under the assumption of constant equation of state $w$. Any dark energy model which is based on a {\it dynamical} equation of state would require a more detailed analysis.

A particularly interesting class of dark energy models is in fact based on fields with equation of state less than $-1$, producing a \emph{phantom (dark) energy}, which is well compatible with present observational constraints. 

\paragraph{Phantom energy.} The mentioned feature of phantom energy compared to other field-theoretic models with $w>-1$, i.e. that its energy density increases as the universe volume grows, is the root of various difficulties in constructing a viable field theoretic model of phantom energy. In fact, $w<-1$ requires negative kinetic energy and leads to a violation of various energy conditions \cite{Caldwell:1999ew,Carroll:2003st,Vikman:2004dc}. The negative kinetic energy is also unbounded from below, and straightforward introductions of a regularizing cutoff would lead, in general, to violations of Lorentz symmetry \cite{Cline:2003gs}. 

While these are serious difficulties for such field-theoretic phantom models, phantom energy cannot be ruled out based on cosmological data. On the contrary, several observations {\it favor} an equation of state less than $-1$ \cite{Nesseris:2006er,Shafer:2013pxa,Zhao:2017cud,Wang:2018fng}.  In addition, it has been recently shown that the existence of phantom energy may alleviate the $H_0$ tension \cite{DiValentino:2020naf,Alestas:2020mvb}, i.e. the fact that the value of the Hubble parameter when estimated from local experiments \cite{Riess:2019cxk} is larger that what is deduced from CMB data \cite{Aghanim:2018eyx}. 

Therefore we seem to be facing a situation in which a phantom-like evolution of the observed (late) universe struggles to find a compelling theoretical description. From our quantum gravity viewpoint, based on a formalism in which spacetime is naturally seen as emergent, the difficulties of a formulation of phantom energy in terms of a field theory framework is not particularly worrying. We expect the whole background cosmological dynamics, including its large-scale features, to be determined by the underlying quantum gravity dynamics, and no fundamental phantom field needs to be part of the story. On the other hand, our task is first of all to match cosmological observations, a difficult challenge for all fundamental quantum gravity approaches, and for this aim an effective phantom dark energy would be suitable. Indeed, we will show in the following how phantom-like dark energy can emerge from our GFT condensate model.

For completeness, we mention that one can also tackle the phantom energy problem in the context of modified gravity theories. That is, one can attribute the accelerated expansion of the universe to a modification of the underlying gravitational dynamics, with respect to GR, rather than to new exotic matter components, for example as a $f(R)$ theory \cite{Nojiri:2006be}. In such a way, one can bypass the difficulties of constructing a well-defined matter field theory of phantom energy. This second approach is much closer in spirit to the one we take within our quantum gravity framework, and the emergent cosmological dynamics we extract from the fundamental quantum dynamics of \lq spacetime constituents\rq\  could in principle be recast also in terms of some effective modified gravity theory.

\paragraph{Big Rip singularity.} Quantum gravity effects can change the evolution of any matter content dramatically. For example, in the early universe, even ordinary matter with $w>-1$ can have phantom like behaviour due to discreteness of quantum geometry \cite{Singh:2005km}. And at late times, quantum gravity effects can dissolve the Big Rip singularity in the presence of phantom matter with $w<-1$, as studied in the LQC context \cite{Samart:2007xz} and also in semi-classical analyses \cite{Haro:2011zzb}. Here we will not consider the coupling between quantum gravity and matter, since the accelerated phase with effective equation of state $w<-1$ will emerge from pure quantum gravity in our model. But a Big Rip singularity is avoided due to a non-trivial time dependence of $w$.  
Indeed, as mentioned above, when $w$ is time dependent the evolution of (any effective) $\rho_\psi$ can be rather involved. In particular, if $w$ approaches to $-1$ fast enough, the phantom energy density does not diverge but increases to a constant value, and the Big Rip singularity can be avoided. For example, consider $\rho_\psi$ as a cosmological constant plus some matter component with negative energy density, inversely proportional to the volume \cite{McInnes:2001zw}. This corresponds to a field $\psi$ with $w_\psi<-1$ but that approaches $-1$ at large volume, so that asymptotically we reach a de Sitter spacetime. This is referred to as \emph{phantom analogues of de Sitter space} in \cite{McInnes:2001zw}. In section \ref{sec:twomodes} and section \ref{sec:bigrip} we will see how exactly this kind of behaviour emerges from our quantum gravity model.

\subsection{$w$ from single-mode GFT condensates}
Before moving on to our new analysis of GFT cosmological dynamics in the presence of interactions, let us summarize earlier work on this issue. The result of \cite{deCesare:2016rsf} is equivalent to a study of the behaviour of the effective $w$ under the assumption that a single GFT field mode $j$ contributes to the dynamics. While the asymptotic dominance of a single mode as the universe expands is expected also in the general case, the presence of other modes changes the way in which $w$ approaches the asymptotic value, which is of important physical relevance, as we explained, on top of making the dynamics much richer in any intermediate regime. But the analysis in \cite{deCesare:2016rsf} is already important to show how GFT interactions can have very interesting consequences on the emergent cosmological dynamics, as we now discuss.

With only one $j$ mode, using \eqref{eq:vpsquare} and \eqref{eq:vpp} in the definition \eqref{eq:effwdef} we have
\iea{
  w&=& \frac{-3Q^2+4E\rho^2+m^2\rho^4+\left(1-\frac{4}{n}\right)\lambda\rho^{n+2}+\left(1-\frac{4}{n'}\right)\mu\rho^{n'+2}}{-Q^2+2E\rho^2+m^2\rho^4-\frac{2}{n}\lambda\rho^{n+2}-\frac{2}{n'}\mu\rho^{n'+2}},
}
where we dropped the subscript denoting different modes $j$ for simplicity. Furthermore, for a single mode the total volume $V\propto\rho^2$, and we can get the evolution $w=w(V)$ even without solving the equation of motion. This greatly simplifies the analysis. 

\paragraph{Early time acceleration in the free case.} At early times, the module $\rho$ of the condensate is small, therefore the interaction terms can be ignored. Here we set $\lambda=\mu=0$, then $w$ is simply
\ieas{
  w=\frac{-3Q^2+4E\rho^2+m^2\rho^4}{-Q^2+2E\rho^2+m^2\rho^4} \qquad .
}
At the bounce, the denominator vanishes, $-Q^2+2E\rho^2+m\rho^4=0$, which gives the value of $\rho$ at the bounce
\ieas{
  \rho_b=\frac{1}{m}\sqrt{\sqrt{E^2+m^2Q^2}-E}.
}
Put this back into $w$ we see that the numerator is negative, therefore $w\to-\infty$ near the bounce. This means that right after the bounce the expansion is accelerating, as we expect from a bouncing scenario\footnote{The universe should expand which requires $V'>0$ after the bounce, and at the bounce we have $V_b'=0$, therefore we should also have $V_b''>0$. Since the volume $V_b$ at the bounce is also positive, from the definition \eqref{eq:effwdef} of $w$ we see that $w\to-\infty$ at the bounce is a general feature.}. However, we can show that this accelerating phase ends quickly, i.e., the volume at the end of acceleration is not large compared to the volume at the bounce \cite{deCesare:2016rsf}. The situation is similar even if we consider the contributions from all modes, as we shall see in section \ref{sec:freecondensate}.

It is worth mentioning that even if $w\to-\infty$ at the bounce, we do not run into singularities due to the quick end of the acceleration phase and the fact that the total volume has a minimum value $V_b>0$. To see this we first note that for a single mode (assumed to be mode $j_0$), the total volume can be given by $V=V_{j_0}\rho^2$, therefore the equation of state can be rewritten as
\ieas{
  w=\frac{-3Q^2V_{j_0}^2+4E V_{j_0}V+m^2V^2}{-Q^2V_{j_0}^2+2E V_{j_0}V+m^2V^2} \qquad .
}
Then we substitute this equation of state into the conservation equation \eqref{eq:psiconservation} for the fictitious field $\psi$, we get the solution
\ieas{
  \rho_\psi=\frac{\tilde{\rho}_{\psi\infty}}{V^2}+\frac{2E_0V_{j_0}}{V^3}\frac{\tilde{\rho}_{\psi\infty}}{m^2}-\frac{Q^2V_{j_0}}{V^4} \frac{\tilde{\rho}_{\psi\infty}}{m^2}\qquad ,
}
where $\tilde{\rho}_{\psi\infty}$ is defined such that $\rho_\psi V^2\to \tilde{\rho}_{\psi\infty}$ as the total volume $V\to\infty$. Since the volume $V\geq V_b>0$ is bounded, we see that the energy density $\rho_\psi$ remains finite, and there is no singularities. We also note that at the bounce we have $\rho_\psi(V_b)=0$.

\paragraph{Emergence of the FLRW universe.} As we have seen in section \ref{sec:volumedynamics}, the classical limit emerges already in the free case. It is obtained at large volume, where $\rho$ is also large. At leading order in $1/\rho$, we have $w=1$ is a constant, corresponds to the equation of state of a free massless scalar field, the one we introduced as relational time. In fact, substituting $w=1$ back into its definition \eqref{eq:effwdef}, simple algebraic manipulation shows that
\ieas{
  \frac{V''}{V}-\left(\frac{V'}{V}\right)^2=\frac{VV''-(V)^2}{V^2}=\frac{\dd}{\dd\phi}\left(\frac{V'}{V}\right)=0,
}
hence $V'/V=const$ which characterizes the FLRW equation using the relational language in the presence of a free massless field \cite{Oriti:2016qtz}. 

At the next order of $1/\rho$, we can approximate $w$ as
\iea{
  w=1+\frac{2E}{m^2\rho^2} \quad ,  \label{eq:wfreeasymp}
}
confirming that the effective equation of state approaches $1$ at large volume. Furthermore, for $E>0$, $w$ approaches this asymptotic value from above; see figure \ref{fig:wlnvfreepara31}. This is not the case when we consider more than one mode, as we shall see in section \ref{sec:twomodesfree}.

\paragraph{An emergent inflationary phase from quantum gravity.} The next question is how the single-mode interactions change this picture, in particular concerning the early acceleration after the bounce. As showed in \cite{deCesare:2016rsf}, one can indeed get a long lasting accelerated phase, in contrast to the free condensate. Furthermore, with two interaction terms this acceleration can end properly, and the time that the acceleration lasts can be adjusted by tuning couplings $\lambda$ and $\mu$ \cite{deCesare:2016rsf}. What is missing, however, is a subsequent FLRW phase, which is of course also crucial for a proper cosmological model. Let us see how this behaviour is reflected in the effective equation of state.  Since we assumed that $|\mu|\ll|\lambda|$, there is an intermediate range, where $m^2\rho^4$ and $\mu\rho^{n'+2}$ are both small compared to $\lambda\rho^{n+2}$, and the behaviour of $w$ is determined by the $\lambda$ term. The $\lambda>0$ case will give an additional root of the denominator of $w$, corresponds to the maximum value of $\rho$ and lead to a cyclic universe very quickly after the bounce. Hence we only consider the case $\lambda<0$, where to leading order we have $w=2-n/2$. We see that for $n\geq5$ we have $w<-1/3$ which corresponds to an accelerating phase. In absence of other interactions, this accelerated phase would simply not end. Otherwise, as $\rho$ increases further, the $\mu$ term becomes important compared to the $\lambda$ term. If $\mu>0$, $\rho'$ will vanish again (besides the point of minimal volume reached at the bounce), corresponding to the maximum value of $\rho$ determined by 
\ieas{
  \frac{2}{n}\lambda\rho^{n+2}=\frac{2}{n'}\mu\rho^{n'+2},
}
near which $w\to\infty$. This means the accelerating phase dominated by the $\lambda$ term stops. By adjusting the values of the couplings $\lambda$ and $\mu$ we can make this phase lasts long enough to account the observational constraints \cite{deCesare:2016rsf}. The magenta dash-dotted line in figure \ref{fig:wintbehaviour} shows the behaviour of $w$ when $\mu>0$ and we see that there is a nice inflationary phase with $w=-1/2$. However, as anticipated, this inflationary phase ends when the volume approaches its maximal value, being quickly followed by a contracting phase, with no FLRW phase in between. The important take home message, however, is that interesting large scale cosmological dynamics, like a long lasting inflationary (or more generally, accelerated) phase can be produced purely from fundamental quantum gravity dynamics, without the need of any exotic matter field (here, an inflaton).

\paragraph{Phantom crossing.} Finally, in this simpler single-mode context, we can ask whether anything like a phantom-crossing can also be obtained as a result of the quantum gravity dynamics. 

As we explained above, when $w<-1$ we have phantom energy. For a dynamical  $w$, it is possible for $w$ to change from $w>-1$ to $w<-1$, a phenomenon called \emph{phantom crossing} \cite{Zhang:2009dw}.  In our case, if $\mu<0$, $\rho$ can keep growing until the $\mu$ term dominates, with the asymptotic behaviour of the equation of state given by
\ieas{
  w\to2-\frac{n'}{2}+\left(n'-n\right)\frac{n'\lambda}{2n\mu}\rho^{n-n'} \qquad .
}
Since $n'>n$, we see that for $n\geq5$, we have $~w<-1/3$ as $\rho$ grows and the acceleration does not stop. And in contrast to the $\mu>0$ case, where the volume has a maximum value after which the universe starts to collapse, when $\mu<0$ the total volume can grow forever. Note that $n'>n$ and that both $\lambda$ and $\mu$ are negative, thus we conclude that $w$ approaches its asymptotic value from above. For $n'=6$, we have $w\to-1$, which mimics the behaviour of a cosmological constant. Since $w$ approaches this value from above, we have $w>-1$ after the end of early accelerating phase (which is dominated by the free parameters of the condensate). We conclude that for a single mode with $n'\leq6$, $w$ cannot cross the phantom divide $w=-1$. This is illustrated in figure \ref{fig:wintbehaviour} by the red dashed line. 

On the other hand, for $n'>6$, the asymptotic value of $w$ would be less than $-1$, so phantom crossing is possible. But now the energy density of the fictitious field $\psi$ with effective equation of state $w$ will diverge as the volume of the universe grows. When the volume is large enough, this energy density would produce a Big Rip singularity \cite{Caldwell:2003vq}. In section \ref{sec:lateacc}, we will show that when we consider two modes, we can get an equation of state $w$ that crosses the phantom divide, and that, instead of a Big Rip singularity, the \emph{phantom analogues of de Sitter space} \cite{McInnes:2001zw} is obtained.

\section{Acceleration in early time} \label{sec:freecondensate}
We now start analyzing our emergent cosmological dynamics, in the case in which GFT interactions are taken into account and two spin modes contribute to it. We focus first on the early universe dynamics, right after the bounce, to see how the presence of two spin modes modifies the results obtained in \cite{deCesare:2016rsf}.

In the last section, we have seen that for a single mode, the universe undergoes an accelerated expansion for a very short period after the bounce. But for the early universe, the volume is small, and in these conditions we have no reason to expect one mode to dominate over the others, so we should consider the contributions from several modes into account. Besides, the smallness of the condensate density $\rho_j$ means that the dynamics of each spin mode is dominated by the free part of the dynamics. Hence, we can consider the free condensate with $\lambda_j=\mu_j=0$ for all $j$. 

\subsection{Accelerated expansion in the free condensate}
We require that $\rho_j'\geq0$ in the region we considered, and then the condition $V'=0$ for the volume at the bounce corresponds to requiring $\rho_j'=0,~\forall j$. The value of $\rho_j$ at the bounce can be obtained by solving the equation $\rho_j'=0$, where $\rho_j'$ is obtained from the definition \eqref{eq:defej} of the GFT \lq\lq energy\rq\rq\  $E_j$ as 
\iea{
  \rho_j'(\phi)&=&\frac{1}{\rho_j}\sqrt{2E_j\rho_j^2-Q_j^2+m_j^2\rho_j^4-\frac{2}{n_j}\lambda_j\rho_j^{n_j+2}-\frac{2}{n'_j}\mu_j\rho^{n'_j+2}} \qquad. \label{eq:rhoprimelambdamu}
}
In the free case, at the bounce we have
\ieas{
  \rho_{bj} =\frac{1}{m_j}\sqrt{\sqrt{E_j^2+m_j^2Q_j^2}-E_j} \qquad.
}
Given the initial value $\rho_j(0)=\rho_{bj}$, the differential equation \eqref{eq:rhoprimelambdamu} can be solved \cite{deCesare:2016axk,Gielen:2016uft} to give
\iea{
  \rho_j(\phi)=\frac{1}{m_j}\sqrt{\sqrt{E_j^2+m_j^2Q_j^2}\cosh(2m_j\phi)-E_j} \qquad . \label{eq:rhosolfree}
}
Then the total volume \eqref{eq:totalVcondensate} becomes 
\iea{
  V&=& \sum_jV_j\rho_j^2= \sum_j\frac{V_j\sqrt{E_j^2+m_j^2Q_j^2}}{m_j^2}\cosh(2m_j\phi)-\sum_j\frac{V_jE_j}{m_j^2} \qquad. \label{eq:freeV}
}
At the bounce $\phi=0$, therefore the volume $V_b$ is simply $V=c_1-c_2$, where $c_1$ and $c_2$ are given by
\iea{
  c_1=\sum_j\frac{V_j\sqrt{E_j^2+m_j^2Q_j^2}}{m_j^2},~c_2=\sum_j\frac{V_jE_j}{m_j^2} \qquad. \label{eq:c1c2def}
}
We can see that $c_1>c_2>0$.

The volume should be convergent, in the sense that $V$ is finite at any given relational time $\phi$. In appendix \ref{sec:convergentV} we show that this is equivalent to the requirement that $\displaystyle \sum_j\frac{V_j}{m_j^2}\sqrt{E_j^2+m_j^2Q_j^2}$ converges and all the $m_j$'s are bounded. A direct consequence is that at sufficiently large $\phi$, the volume is dominated by the mode with the largest value of $m_j=m$. This largest value defines, in this regime, the effective Newton's constant $m^2=3\pi G$, and the dynamics reduces to the standard Friedmann equation with the matter content given by the free massless scalar field \cite{Oriti:2016qtz}. There are general arguments suggest that $m_j$ is monotonically decreasing with $j$, so that, at large volume, it is the smallest spin mode that eventually dominates \cite{Gielen:2016uft}.

\subsection{Upper bound of the number of e-folds}

Now we are ready to check if the inclusion of all modes can make the acceleration phase after the bounce last long enough to be of phenomenological significance as a quantum gravity-induced inflation, even in the free case. 

For simplicity, we introduce a function $P(\phi)$ to characterize the acceleration
\iea{
  P(\phi)=-\frac{(V')^2}{2}\left(w+\frac{1}{3}\right) \qquad, \label{eq:Pdef}
}
taking in this free case the form
\iea{
  P(\phi)&=&  \sum_j4V_j\sqrt{E_j^2+m_j^2Q_j^2}\cosh(2m_j\phi) \sum_k\frac{V_k}{m_k^2}\left[\sqrt{E_k^2+m_k^2Q_k^2}\cosh(2m_k\phi)-E_k\right] \nonumber  \\
  &&-\frac{5}{3}\sum_j\frac{2V_j}{m_j}\sqrt{E_j^2+m_j^2Q_j^2}\sinh(2m_j\phi)\sum_k\frac{2V_k}{m_k}\sqrt{E_k^2+m_k^2Q_k^2}\sinh(m_k\phi) \qquad. \label{eq:Pfree}
}
The accelerating expansion requires $w<-1/3$, i.e. $P(\phi)>0$, while the decelerating phase corresponds to $P(\phi)<0$.

At the bounce, where $V'=0$, we have simply
\iea{
  P(0)=\sum_j4V_j\sqrt{E_j^2+m_j^2Q_j^2}\left(c_1-c_2\right)>0,
}
with $c_1$ and $c_2$ defined in \eqref{eq:c1c2def}, while for large $\phi$, the volume is dominated by a single mode, and equation \eqref{eq:wfreeasymp} tells us that $w\to1$ when volume is large. This implies $P(\phi)<0$ at large volume. Therefore, there is a point where $P(\phi)=0$ and the accelerating expansion stops. We now identify this point and show that the accelerating phase until then can not be long enough. More precisely, we get an upper bound on the ratio $V_e/V_b$, where $V_e$ is the volume when acceleration ends, and $V_b=c_1-c_2$ is the volume at the bounce. 

The time $\phi_e$ where the accelerating phase ends is determined by the requirement $P(\phi_e)=0$. This equation is quite hard to solve for general $m_j$'s. If the acceleration is long lasting, $\phi_e$ would be large, and around this point $P(\phi)$ changes quickly. Therefore we can introduce an approximated quantity $P_m(\phi)$, obtained by replacing $\cosh(2m_j\phi)$ and $\sinh(2m_j\phi)$ in \eqref{eq:Pfree} with $\cosh(2m\phi)$ and $\sinh(2m\phi)$ respectively, where $m$ is the maximum value of $m_j$'s. We can write $P_m(\phi)$ as 
\ieas{
  P_m(\phi)&=&- \frac{4m}{3}\left[\cosh^2(2\sqrt{m}\phi)\left(5c_1'^2-3c_1c_1''\right)+\cosh(2\sqrt{m}\phi)(3c_1''c_2)-5c_1'^2\right],
}
with $c_1$ and $c_2$ are given by equation \eqref{eq:c1c2def} and the two new constants $c_1'$ and $c_2'$ are
\iea{
  c_1'&=& \sum_j\frac{V_j}{mm_j}\sqrt{E_j^2+m_j^2Q_j^2},~c_1''=\sum_j\frac{V_j}{m^2}\sqrt{E_j^2+m_j^2Q_j^2} \qquad. \label{eq:c1pc1ppdef}
}
We see that $c_1>c_1'>c_1''>0$. The equation $P_m(\phi)=0$ has a root $\phi_m$ and one has
\iea{
  \cosh(2m\phi_m)&=& \frac{-3c_1''c_2+\sqrt{9c_1''c_2^2+20c_1'(5c_1'^2-3c_1c_1'')}}{2(5c_1'^2-3c_1c_1'')} \qquad .
}

Since $P(\phi)$ changes quickly near $\phi_e$, we have approximately $\phi_e\approx\phi_m$, which in turn leads to $V_e=V(\phi_e)\approx V(\phi_m)<V_m(\phi_m)$. Here we define $V_m$ similarly as $P_m$, i.e., replacing $\cosh(2m_j\phi)$ in the volume \eqref{eq:freeV} with $\cosh(2m\phi)$, and therefore, at $\phi=\phi_m$ we have
\ieas{
  V_m(\phi_m)=c_1\cosh(2m\phi_m)-c_2.
}
Then the ratio between volume at the end of acceleration and the volume at the bounce satisfies
\iea{
  \frac{V_e}{V_b}<\frac{V_m(\phi_m)}{V_b}=\frac{-3c_1c_1''c_2+c_1\sqrt{9c_1''c_2^2+20c_1'(5c_1'^2-3c_1c_1'')}}{2(5c_1'^2-3c_1c_1'')(c_1-c_2)}-\frac{c_2}{c_1-c_2}.
}
with $c_1$ and $c_2$ are given by equation \eqref{eq:c1c2def}. Under the conditions $c_1>c_2>0$ and $c_1>c_1'>c_1''>0$, $V_{m}(\phi_m)/V_b$ has a maximum value 
\iea{
  \left.\frac{V_{m}(\phi_m)}{V_b}\right|_{max}=1+\frac{c_1}{c_2}+\frac{c_1}{c_2}\sqrt{\frac{c_1+c_2}{c_1-c_2}} \qquad . \nonumber
}
Therefore, the original volume ratio with $j$ dependent $m_j$ has the upper bound
\iea{
  \frac{V_e}{V_b}<1+\frac{c_1}{c_2}+\frac{c_1}{c_2}\sqrt{\frac{c_1+c_2}{c_1-c_2}}, \label{eq:vevbbound}
}
with $c_1$ and $c_2$ are defined in equation \eqref{eq:c1c2def}.

The bound goes to infinity when $\frac{c_2}{c_1}\to0$ or $\frac{c_2}{c_1}\to1$. However, since the total volume $V$ should be finite, both of $c_1$ and $c_2$ should be finite. Then, using their definition, we see that $\frac{c_2}{c_1}\to0$ would require $E_j\to0$ for all $j$ while $\frac{c_2}{c_1}\to1$ would require $Q_j^2\to0$ ($m_j$ cannot vanish otherwise $c_1$ and $c_2$ would diverge) for all $j$. Therefore, for general configurations corresponding to non-vanishing $Q_j$ and $E_j$ for some $j$, the bound on the number of e-folds would not be large. While for vanishing $Q_j$ and $E_j$ we need to find a different bound to reach a reliable conclusion, it is clear that this would correspond to a rather special case, thus of limited interest, especially in a phenomenological setting like this.

We conclude that the expansion of the universe becomes decelerating quickly after the bounce, confirming in this more general setting the results of \cite{deCesare:2016rsf}. 

We emphasize that this initial accelerating expansion is in fact a general feature of a bouncing universe, not necessarily linked to any inflationary-like scenario. Inflation as usually understood should instead start later, during the radiation dominating phase \cite{Martin:2019zia}. Such later inflationary acceleration can indeed be reproduced as it has been shown in the previous section, recalling the results of \cite{deCesare:2016rsf}, 
 when accounting for GFT interactions in our condensate. As we discussed, however, single-mode interactions which are strong enough to be relevant shortly after the bounce, and before a FLRW phase produced by the free GFT dynamics, end up preventing that such a FLRW phase is realized after the inflationary one, in contrast to a physically viable cosmological model. One may wonder if the contribution from multiple modes changes this picture. A moment of reflection, together with the analysis we present in the next section, would convince that this may only be possible in the presence of somewhat extreme fine-tuning of parameters and a very special behaviour of the condensate density, since in practice it would require that the contributions from the two interaction terms for the two modes approximately cancel for a long enough period of relational time, after the inflationary phase, so to effective reproduce the free dynamics and its FLRW phase. A situation of this type, even if possible in principle, would be of little interest, unless somehow governed by some symmetry principle or some other generic feature of the underlying quantum gravity model. Lacking this, we do not consider it further in the following.

 We discuss instead in detail the role of GFT interactions in producing an accelerated expansion at even later times, in the next section. The important point to stress here is that, as long as the interaction couplings are small compared to \lq mass\rq\  term $m_j$, the behaviour of condensates can be well approximated by free solutions. Therefore, a very short-lived accelerated expansion after the bounce followed by a a decelerating phase remains a general feature even in the presence of interactions. We are going to use this feature to ensure that, whatever the detailed late time evolution of the universe in our model is, an extended FLRW phase can be realized, before quantum gravity interactions become relevant, as required by observations.

\subsection{Equation of state after the end of acceleration} \label{sec:twomodesfree}
\begin{figure}[htp]
	  \begin{center}\includegraphics[width=0.45\textwidth]{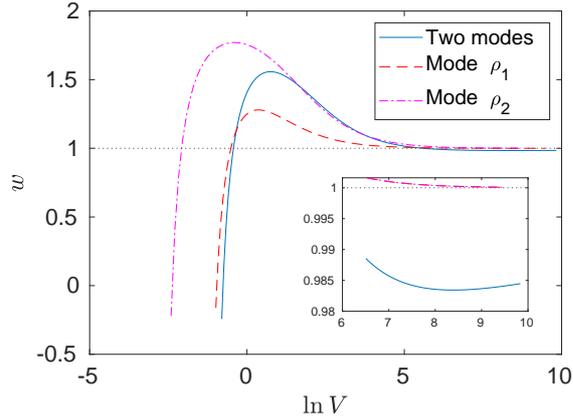}\end{center}
	  \caption{The behaviour of $w$ for different modes in the free case. Blue solid line considers contributions from both $\rho_1$ and $\rho_2$. Red dashed line shows the evolution of $w$ of mode $\rho_1$, while magenta dash-dotted line shows $\rho_2$ case. For the convenience we also plotted the constant $w=1$ using black dotted line. In the little box we showed the finer structure at large volume. We see that $w<1$ in the two modes case, while $w>1$ in both the single mode case $\rho_1$ and $\rho_2$. At large volume the value of $w$ for $\rho_1$ and $\rho_2$ are so close that they can't be distinguished from each other in the plot. Parameters are $V_1=1/3,m_1^2=3,~E_1=5,~Q_1^2=9,V_2=1/2,~m_2^2=2,~E_2=9,~Q_2^2=2.25$.}
	  \label{fig:wlnvfreepara31}
\end{figure}

More precisely, after the end of the post-bounce acceleration, the expansion itself does not stop and the volume of universe keeps growing. According to the free solution \eqref{eq:rhosolfree}, for large $\phi$ the module $\rho_j$ increases exponentially. Therefore the mode with largest $m_j$ dominates quickly as the volume growing, which means the equation of state will soon be dominated by this single mode as well. As we have already discussed, $w$ will have the asymptotic value $w=1$ as in the single mode case, corresponding to the equation of state of the free massless scalar field that we are using as relational time. However, the inclusion of other modes changes the precise way in which $w$ approaches to the asymptotic value. Take the two-modes case as an example (for simplicity, we write $\rho_{1,2}\equiv \rho_{j_1,j_2}$ etc). Assuming $m_1>m_2$ and hence at large volume we have $\rho_1>\rho_2$, then $w$ can be expanded as
\ieas{
  w\to1+\frac{2V_2\rho_2^2}{V_1\rho_1^2}\left(2\sqrt{\frac{m_2}{m1}}-1-\frac{m_2}{m_1}\right) \qquad.
}
Since $2\sqrt{m_1/m_2}<1+m_2/m_1$, we see that $w$ approaches the asymptotic value from below, in contrast with the single mode case. In figure \ref{fig:wlnvfreepara31} we compare the behaviour of $w$ in the two-modes case and in single-mode case. At small volume near the bounce, $w<0$ and its absolute value is large; this corresponds to large acceleration right after the bounce. With the increase in volume, $w$ grows quickly and becomes larger than $-1/3$ soon, where the accelerated expansion stops. Then $w$ keeps growing and reaches its maximum value, after which $w$ starts to decrease. This behaviour is true for both the two-modes and single-mode cases. As the volume grows further, the evolution of $w$ starts to differ in the two cases. In the two-modes case, $w$ has a minimum value, which is smaller than $1$, after which $w$ starts to increase again, and reaches $w=1$ from below. In the single-mode case, instead, there is no local minimum, and $w$ keeps decreasing, and approaches the asymptotic value $w=1$ from above. Something similar will happen in the interacting case. We will see that, for interactions of order $6$, the asymptotic value will be the phantom divide $w=-1$. Therefore at large volume we have $w<-1$ and the phantom divide is crossed.

\section{Late time accelerated expansion} \label{sec:lateacc}
We now turn to the main focus of our analysis, i.e. the emergent cosmological dynamics of interacting multi-mode condensates at late times.

In the last section, we have seen that for a free condensate, the accelerated expansion only lasts for a short while after the bounce. As volume increases, the quantum gravity condensate would then be descried by a FLRW universe filled with a single massless scalar field. For large condensate densities (and thus volume), however, we expect the interactions to be relevant.  

We first discuss how to solve the equation of motion for each mode, at least approximately. Then we extract the asymptotic behaviour of the effective equation of state $w$ in the two-modes case, showing that it is possible for the phantom divide to be crossed, thus producing a phantom-like dark energy purely from quantum gravity effects. In contrast to the single mode case, moreover, the phantom crossing does not lead to a Big Rip singularity. Finally, we also show that it is possible to produce at late times a more involved, if maybe less phenomenologically interesting, combination of inflation-like and phantom-like dark energy in our model.

 \subsection{Large $\rho$ behaviour of the interacting condensate} \label{sec:interactcondensate}

\begin{figure}[htp]
	  \begin{center}\includegraphics[width=0.45\textwidth]{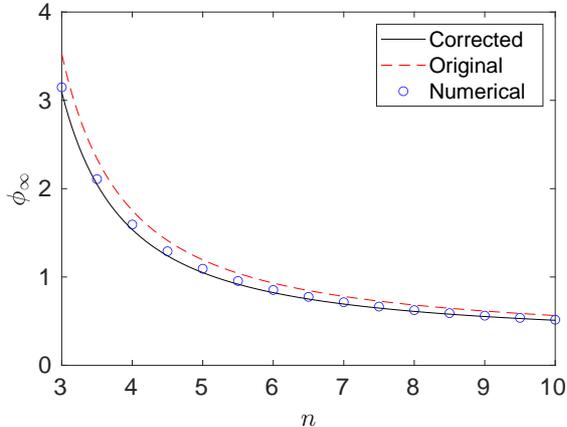}\end{center}
	  \caption{Asymptotic value $\phi_{\infty}$ for different $n$. Black solid line is obtained from equation \eqref{eq:phiinfncor}, the corrected value of $\phi_{\infty}$. Red dashed line is the uncorrected value of $\phi_{\infty}$, given by equation \eqref{eq:phiinfn}. Blue circles shows the numerical results obtained by solving the equation of motion \eqref{eq:rhoprimelambdamu} numerically (with $\mu_j=0$) and set $\rho$ to be large. Parameters are $m^2=2,~E=9,~Q^2=2.25,~\lambda=-0.1$.}
	  \label{fig:phiinfnsolpara31}
\end{figure}

 With interactions being included, the equation \eqref{eq:rhoprimelambdamu} is much harder to solve, and in general the solution cannot be written in close analytic form. Nevertheless, under our assumption that $|\mu_j|\ll|\lambda_j|\ll m_j^2$, the equation of motion can be solved piece-wisely. For simplicity, we first assume $\mu_j=0$; then for $\lambda_j<0$ and $\rho_j$ is large, the equation \eqref{eq:rhoprimelambdamu} can be approximated as
\iea{
  \rho_j'(\phi)=\sqrt{\frac{-2\lambda_j}{n_j}}\rho(\phi_j)^{\frac{n_j}{2}} \qquad . \label{eq:rholargediff}
}
This equation can be easily solved and gives
\iea{
  \rho_j(\phi)=\left(\frac{2}{n_j-2}\sqrt{\frac{-2\lambda_j}{n_j}}\right)^{-\frac{2}{n_j-2}}\frac{1}{(\phi_{j\infty}-\phi)^{\frac{2}{n_j-2}}} \quad, \label{eq:rholargesoln}
}
where $\phi_{j\infty}$ is a constant of integration, determined by initial conditions. Its value can be fixed by matching with solutions in the free case \eqref{eq:rhosolfree}. We choose the matching point $\rho_{j0}$ to be where the \lq mass\rq\  term equals to the interaction term, $m_j^2\rho_{j0}^2=-2\lambda_j\rho_{j0}^{n_j}/n_j$, i.e. the point where the two approximations we used to solve the dynamical equation reach their limit of validity. Assuming that the free solution \eqref{eq:rhosolfree} is valid up to $\rho_{j0}$ for each individual $j$, then $\phi_{j0}$ can be determined inverting the solution \eqref{eq:rhosolfree}. Taking $(\phi_{j0},\rho_{j0})$ as an initial condition for the equation \eqref{eq:rholargediff} and then inserting them into the solution \eqref{eq:rholargesoln}, we can get an approximate value of the constant $\phi_{j\infty}$ as
\iea{
  \phi_{j\infty} &=& -\frac{\ln[-\lambda_j/(2m_j^2)]}{(n_j-2)m_j} +\frac{1}{2m_j}\ln\left[\frac{n_j^{\frac{2}{n_j-2}}(2m_j^2)}{\sqrt{E_j^2+m_j^2Q_j^2}}\right]-\frac{\ln2-1}{m_j}\frac{2}{n_j-2} \qquad. \label{eq:phiinfn}
}
Furthermore, the accuracy of our approximate result of $\phi_{j\infty}$ can be improved with the help of exact solutions in special cases. As showed in appendix \ref{sec:phiinfn4}, for $n_j=4$ the equation of motion \eqref{eq:rhoprimelambdamu} can be solved using elliptic functions. Then using the fact that $|\lambda_j|$ is small, an expansion of $\phi_{j\infty}$ can also be obtained. By comparing with the result in \eqref{eq:phiinfn}, we see that an additional term $\frac{\ln2-1}{m_j}\frac{2}{n_j-2}$ should be added, and the corrected form of $\phi_{j\infty}$ becomes
\iea{
  \phi_{j\infty} &=& -\frac{\ln[-\lambda_j/(2m_j^2)]}{(n_j-2)m_j} +\frac{1}{2m_j}\ln\left[\frac{n_j^{\frac{2}{n_j-2}}(2m_j^2)}{\sqrt{E_j^2+m_j^2Q_j^2}}\right]. \label{eq:phiinfncor}
}
We can compare this form of $\phi_{j\infty}$ for a given mode $j$ with its numerical value, obtained by solving the equation of motion \eqref{eq:rhoprimelambdamu} numerically and substituting a large $\rho_j$ (here taken to be $\rho_j=10^8$) into the solution. The result is shown in figure \ref{fig:phiinfnsolpara31}. We see that our formula also works for non-integer $n_j$ and, despite various approximations, the result is quite accurate at the order of $\lambda_j$. For, comparison, we also plot the original $\phi_{\infty}$, given by \eqref{eq:phiinfn} without correction, which shows that the additional term indeed improves the accuracy of our result.

It is clear from equation \eqref{eq:phiinfncor} that, for each mode $j$, the corresponding $\phi_{j\infty}$ is different. Note that $\rho_j(\phi)$ diverges when $\phi=\phi_{j\infty}$, hence the total volume $V=\sum_jV_j\rho_j^2$ will diverge when $\phi$ reaches $\phi_\infty=\min\{\phi_{j\infty}\}$, the smallest one of the different $\phi_{j\infty}$'s corresponding to different modes. Moreover, when $V$ is large enough, the mode with $\phi_{j\infty}=\phi_{\infty}$ will dominate. 

To the leading order of $\lambda_j$, we have
\ieas{
  \frac{\partial \phi_{j\infty}}{\partial m_j}=\frac{\ln[-\lambda_j/(2m_j^2)]}{(n_j-2)m_j^2} \qquad.
}
For small $|\lambda_j|$, this derivative is less than $0$, thus large $m_j$ will give small $\phi_{j\infty}$. Therefore, also with interactions the condensate dynamics tends to be dominated by the mode with largest $m_j$, which in general corresponds to small-$j$ modes as in the free case. We note here, anticipating the discussion in section \ref{sec:bigrip}, that the volume divergence at finite relational time $\phi$ does not necessarily imply the existence of Big Rip singularity. In fact, if we consider the fictitious field $\psi$ with equation of state equals to $w$, then for $n\leq 6$ its energy density $\rho_\psi$ will remain finite for $V\to\infty$; see section \ref{sec:bigrip} for details.

We emphasize that the solution \eqref{eq:rholargesoln} only works for negative couplings. In fact, if we add another interaction term $\mu_j>0$, even under the assumption that $|\mu_j|\ll|\lambda_j|$, so that  the contribution of $\mu_j$ to the value of $\phi_{j\infty}$ can be ignored, the behaviour of $\rho_j$ at late times changes considerably. Explicitly, for $\mu_j>0$, from equation \eqref{eq:rhoprimelambdamu} we see that, besides the bounce, $\rho'_j(\phi)=0$ has an additional solution for some large $\rho_j$, determined by $\displaystyle \frac{2}{n}\lambda\rho^{n+2}=\frac{2}{n'}\mu\rho^{n'+2}$,  which corresponds to the maximum value of $\rho_j$ (and thus of the volume) at later times. After that, to ensure $\rho_j'$ is real, we should require that $\rho_j$ starts to decrease, and it leads to a periodic evolution of $\rho_j$ and thus a cyclic universe (as in \cite{deCesare:2016rsf}). Since $|\mu_j|\ll|\lambda_j|$, we can take the value of $\phi$ approximately as $\phi\approx \phi_{j\infty}$ where $\rho_j$ first reaches its maximum. Therefore, in the case with $\mu_j>0$, instead of being the largest value that $\phi$ can reach (as in the single interaction case), $\phi_{j\infty}$ now should be regarded as a half-period in the evolution of $\rho_j$, indicating that $\rho_j$ actually starts to decrease for $\phi>\phi_j$. On the other hand, for $\mu_j<0$, $\rho_j$ can keep growing until $\phi$ reaches $\phi_{j\infty}$ where $\rho_j$ diverges. 

We will see in the next section how the combination of two modes with opposite sign of $\mu_j$ makes it possible for the effective equation of state to cross the phantom divide $w=-1$.

\subsection{Phantom crossing in the two-modes case} \label{sec:twomodes}

In this section we consider how the presence of two interacting modes, each with an individual contribution to the cosmological dynamics of the type we have illustrated above, can shape it in very interesting ways at late times.

For simplicity, we use $\rho_{1,2}$ to indicate $\rho_{j_1,j_2}$ and similarly for other parameters. Although in previous sections we have seen that at sufficiently large volume there will be only one mode dominating also in the interacting case, we will see that the inclusion of a second mode does change the behaviour of the effective equation of state $w$, and in particular how the asymptotic value is approached, which is of direct cosmological relevance. 

To begin with, we consider the case in which two modes both have a single interaction term, i.e. we set $\mu_1=\mu_2=0$. Since the coupling $\lambda_1$ and $\lambda_2$ are small, $w$ will be dominated by the free part of condensate at small volume, and it will approach $w=1$ from below as volume grows. This is the needed FLRW universe of the standard cosmological model, reached after the phase close to the big bang, here replaced by a quantum bounce. When the volume becomes larger still, the interaction term for both modes increasingly contributes to the condensate dynamics, until, for large enough values (of $\rho_j$ and thus of the volume), $w$ will be dominated by the interaction terms instead. If we further assume that $n_1=n_2=n$, considering only interaction terms in the expression for $w$ would suggest that $w$ only depends on the ratio $r=\rho_2/\rho_1$ (as it was the case also in the free case discussed above), and we have
\iea{
  w&=& 3-\frac{(2+n)(V_1+r^2V_2)(V_1\lambda_1+r^nV_2\lambda_2)}{2\left(V_1^2\lambda_1+r^{2+n}V_2^2\lambda_2-2r^{1+\frac{n}{2}}V_1V_2\sqrt{\lambda_1\lambda_2}\right)}\,= \nonumber  \\
  &=& 2-\frac{n}{2}-\left(\frac{n}{2}+1\right)\frac{V_1V_2r^2\left(r^{n/2-1}-\sqrt{\lambda_1/\lambda_2}\right)^2}{\left(\sqrt{\lambda_1/\lambda_2}V_1+V_2r^{n/2+1}\right)^2} \qquad . \label{eq:wrrho2rho1}
}
Since the parameters are all real and both couplings $\lambda_1$ and $\lambda_2$ are assumed to be negative, we see that $\displaystyle w\leq2-\frac{n}{2}$. Recall that when the volume is large, one of the two modes will dominate over the other, and then we have $r\to0$ or $r\to\infty$. In either case $w$ will approach $\displaystyle 2-\frac{n}{2}$ from below, in contrast with the single mode case discussed in section \ref{sec:effwsingle}.

There is a special case where $\displaystyle r=\left(\frac{\lambda_1}{\lambda_2}\right)^{\frac{n}{4}-\frac{1}{2}}$, then, within the approximation we have made, we see that $w=2-n/2$ is also a constant. From our solution \eqref{eq:rholargesoln} for each mode at large volume, we see that this indeed happens when $\phi_{1\infty}=\phi_{2\infty}$. In fact, when $\rho_2=r\rho_1$ is proportional to $\rho_1$, we have $V=V_1\rho_1^2+V_2\rho_2^2=(V_1+r^2V_2)\rho_1^2$, which is the same as the single mode case with a modified $\tilde{V}_1=V_1+r^2V_2$. And therefore the equation of state is the same as in the single mode case, which indeed approaches the asymptotic value from above. 

\begin{figure}[htp]
	  \begin{center}\includegraphics[width=0.45\textwidth]{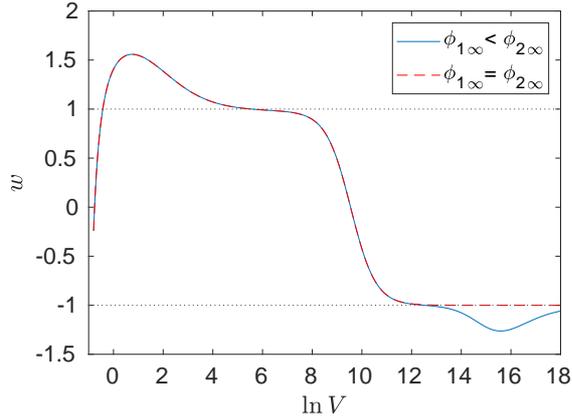}\end{center}
	  \caption{The behaviour of $w$ in the two modes case, where both modes have only one interaction term. Blue solid line shows the case where $\phi_{1\infty}<\phi_{2\infty}$, while for red dashed line we have $\phi_{1\infty}=\phi_{2\infty}$. Two black dotted lines show $w=1$ and the phantom divide $w=-1$, respectively. Parameters are same as in figure \ref{fig:wlnvfreepara31} with additional ones are $\lambda_1=-10^{-8},~\mu_1=0,~\mu_2=0,~n_1=n_2=6$ and $\lambda_2=-9.5\times10^{-8}$ for $\phi_{1\infty}<\phi_{2\infty}$, $\lambda_2=-9.5725\times10^{-8}$ for $\phi_{1\infty}=\phi_{2\infty}$.}
	  \label{fig:wlnvpara37p38}
\end{figure}

In figure \ref{fig:wlnvpara37p38} we plot the different behaviour of $w$ in the cases $\phi_{1\infty}<\phi_{2\infty}$ and $\phi_{1\infty}=\phi_{2\infty}$ using numerical solutions of equation of motion \eqref{eq:rhojpp} in the single interaction case $\mu_j=0$. 

At small volume, the evolution is dominated by the free parameters, and the two case are identical. 
At a larger volume but when $\phi$ is still away from $\phi_{1\infty}$, the ratio $r=\rho_2(\phi)/\rho_1(\phi)$ changes slowly, and the behaviour of $w$ in the two cases is still almost identical. 
As the volume grows further, $\phi$ approaches to $\phi_{1\infty}$, then in the case $\phi_{1\infty}<\phi_{2\infty}$, $\rho_1$ tends to $\infty$ and grows much fast than $\rho_2$, leads to $r\to0$, and $w$ approaches to the phantom divide $w=-1$ from below. In the same regime, but for $\phi_{1\infty}=\phi_{2\infty}$, we have $\displaystyle r=\lambda_1/\lambda_2$, so the last term in \eqref{eq:wrrho2rho1} vanishes, and $w$ will approach $w=-1$ from above, as in the single mode case.

Now we consider the case $n=6$ and assume that $\phi_{1\infty}<\phi_{2\infty}$. Then at large volume the first mode will dominate and $r\to0$. Expanding $w$ in equation \eqref{eq:wrrho2rho1} with respect to $r$ gives simply
\ieas{
  w=-1-\frac{4V_2}{V_1}r^2=-1-\frac{4V_2}{V_1}\frac{\rho_2(\phi)^2}{\rho_1(\phi)^2} \qquad .
}
Therefore, when $n=6$ the phantom divide $w=-1$ can be crossed at large volume and the corresponding effective field $\psi$ behaves just like a phantom energy, whose energy density increases as the volume of universe grows. 

This is our main result, showing how a phantom-like dark energy dynamics at late times can be produced, under rather general conditions (albeit in a simplified model, and of course in a specific regime of the full theory) purely from quantum gravity effects, i.e. as an effective description of the underlying quantum dynamics of spacetime constituents. 

One may then worry about whether this effective phantom energy, like in many field theoretic models, leads to a Big Rip singularity at later times also in our model. We will discuss this issue in the next section, showing that the effective energy density $\rho_{\psi}$, {\it defined} from the equation of state $w$, remains bounded in our model, tending towards to a finite value at asymptotically large volumes. To see this, we need some further approximation for the equation of state $w$, which we anticipate here. 

Since $\phi_{1\infty}<\phi_{2\infty}$, and for large volume we have $\phi\to\phi_{1\infty}$, we see that $\rho_2$ is nearly a constant given by $\rho_2(\phi_{1\infty})$. Using the solution \eqref{eq:rholargesoln}, we get
\ieas{
  \rho_2(\phi_{1\infty})= \left(\frac{1}{2}\sqrt{\frac{-\lambda_2}{3}}\right)^{-\frac{1}{2}}\frac{1}{(\phi_{2\infty}-\phi_{1\infty})^{\frac{1}{2}}} \qquad.
}
Furthermore, when $\phi\to\phi_{1\infty}$ the first mode $\rho_1$ would be much larger than $\rho_2$, hence in computing the total volume we can ignore $\rho_2$ and let $V=V_1\rho_1^2$. Inserting this approximate expression back in the expression for $w$, we get
\iea{
  w&=& -1-\frac{b}{V}, \label{eq:wlargev}
}
where $\displaystyle b=4V_2\rho_2(\phi_{1\infty})$ is a constant. Notice again that $b>0$, thus we have $w<-1$, and the phantom divide $w=-1$ is being crossed. 

\subsection{The Big Rip singularity} \label{sec:bigrip}

We pointed out that in the presence of interactions $\rho_j$ and hence the volume will diverge at finite relation time $\phi_{\infty}=\min\{\phi_{j\infty}\}$. Now we show why this does not necessarily mean that a Big Rip singularity is reached. Also, the phantom crossing $w<-1$ would raise the same worry, but, as we already mentioned, only for constant equation of state. We now see why such singularity does not occur in our setting. 

Consider the fictitious field $\psi$ we introduced with equation of state equals to $w$. Its energy density $\rho_\psi$, {\it defined} by the equation of state itself, satisfies the conservation equation \eqref{eq:psiconservation}. We can then substitute for $w$ the approximate expression \eqref{eq:wlargev}, to get
%
\ieas{
  \frac{\dd \rho_\psi}{\dd V}-\frac{b\rho_\psi}{V^2}=0.
}
We can then solve for $\rho_\psi$ at large volume as
\iea{
  \rho_\psi=\rho_{\psi0}\ee^{-\frac{b}{V}}\approx\rho_{\psi0}-\frac{\rho_{\psi0}b}{V} \quad ,
}
where $\rho_{\psi0}$ is a constant of integration, representing the asymptotic value of $\rho_\psi$ when $V\to\infty$. 

Thus we see that we obtain a constant asymptotic value for the energy density, which has the same effect as a cosmological constant. Therefore our model leads to a de Sitter spacetime asymptotically, with no Big Rip singularity. In fact, our model effectively belongs to the class of models considered in \cite{McInnes:2001zw}, where the Big Rip singularity is avoided even in presence of phantom matter by assuming that $\rho_\psi$ can be obtained as a constant part plus some matter with negative energy density. 
Exactly this type of scenario is reproduced from the fundamental quantum gravity dynamics. 

Let us stress that, in order to obtain a de Sitter spacetime asymptotically, the requirement that $w$ approaches to the phantom divide $w=-1$ at large volume is a necessary but not sufficient condition. We need also that $w$ approaches to $w=-1$ fast enough, as it happens naturally in our case. 
To see this, suppose that, when volume $V$ is larger than some given $V_0$, the equation of state can be approximated by 
\ieas{
  w=-1-\frac{b}{\ln (V/V_0)} \qquad .
}
Substituting this into the conservation equation \eqref{eq:psiconservation}, the evolution of the phantom energy density $\rho_\psi$ now reads
\ieas{
  \rho_\psi=\rho_{\psi0}\left[\ln (V/V_0)\right]^b,
}
where $\rho_{\psi0}$ is again a constant, now given by the energy density at volume $V=\ee V_0$. In this case $\rho_\psi$ diverges when $V\to\infty$, and we reach a Big Rip singularity rather than the asymptotically de Sitter spacetime.

\subsection{More involved late-time behaviour: combined inflation-like and phantom-like acceleration} \label{sec:infphantom}

We have seen that we can reproduce naturally the late time acceleration behaviour of our observed universe with a single interaction term for each mode. We also have reasons to expect that the late-time cosmological dynamics is dominated by a single interaction (that of the highest order, if more than one is allowed with comparable weights by the parameters of the model). Thus, we can claim some degree of generality for our main results. 

However, it is interesting to ask how the late-time dynamics, after a FLRW phase, is affected by the presence of multiple interactions, for each mode. This could be relevant for further cosmological applications, but it also has purely theoretical motivations. For example, although $n=6$ interactions are needed to reproduce phantom crossing, most quantum geometric TGFT models include $n=5$ interactions because they come from the simplicial construction of their (lattice gravity and spin foam) amplitudes \cite{Oriti:2011jm}. 

So, we conclude our present analysis by considering briefly the case in which two spin modes both have two interactions, with the new couplings being $\mu_1$ and $\mu_2$. 

When both $\mu_1$ and $\mu_2$ are less than $0$, both modes would produce a divergent condensate density eventually and lead to a similar result as the previous single interaction case. On the other hand, when both $\mu_1$ and $\mu_2$ are positive, there would be a turning point for the condensate density for each mode, after which $\rho_j$ starts to decrease, and the corresponding universe would become cyclic, as in the single mode case. The more interesting case, therefore, is when $\mu_1$ and $\mu_2$ have different signs. 

We assume then that $\mu_1<0$ while $\mu_2>0$. 
As shown in \cite{deCesare:2016rsf}, the mode $\rho_2$ alone can lead to a long-lasting inflationary-like phase. Now, with and additional mode $\rho_1$, we can have both the late time phantom-like acceleration as well as an inflationary-like phase before it.

With two interactions, we have three different cases according to the relative magnitude between $\phi_{1\infty}$ and $\phi_{2\infty}$. Since $\mu_2>0$, $\phi_{2\infty}$ would be the half-period of the $\rho_2$ mode rather than the maximum value that $\phi$ can reach as $\rho_2\to\infty$. For $\phi_{1\infty}<\phi_{2\infty}$, the $\rho_1$ mode would increase faster than the $\rho_2$ mode, and dominate before inflation can end, leading to a similar dynamics as in the single interaction case. On the other hand, for $\phi_{1\infty}>\phi_{2\infty}$, $\rho_2$ will reach its maximum value before $\rho_1$ diverges. For large volume, but with $\phi<\phi_{2\infty}$, the $\rho_2$ mode would dominate and hence inflation can end. But since near $\rho_{2\infty}$, $\rho_2$ decreases very quickly, the total volume will also decrease for a while and then increase again when the $\rho_1$ mode takes over. Let us look at the resulting dynamics in more detail, considering the case where $\phi_{1\infty}=\phi_{2\infty}$ and assuming $n_1=n_2=5,~n_1'=n_2'=6$. 

Since the absolute value of the couplings $|\mu_{1,2}|$ is much less than $|\lambda_{1,2}|$, there would still be a region where the $\lambda$ interaction terms dominate. Furthermore, we can also ignore the influence of $\mu$ terms on the value of $\phi_{j\infty}$, and the solution of each modes can still be given by equation \eqref{eq:rholargesoln} in this region, with $\phi_{1\infty}=\phi_{2\infty}$. Then, as we discussed, in such case the ratio $\rho_1/\rho_2$ becomes a constant and the contribution from two modes cancels, leaves a constant equation of state $\displaystyle w=-\frac{1}{2}$ in this region, corresponding to an inflationary-like phase. 

As the volume increases, the $\mu$ terms become important. In this region, the equation of state $w$ will increase first, and inflation will end after $w>-1/3$. Afterwards, $w$ decreases again to cross the phantom divide $w=-1$. At very large volume, the equation of state can still be approximated by $w=-1-b/V$, only this time with the constant $b$ given by $\displaystyle b=\frac{144}{25}\frac{V_2\lambda_2^2}{\mu_2^2}$, which can be determined with the parameters of the second mode only. 

\begin{figure}[htp]
  \centering
   \subfigure[~Behaviour of $w$ in the interacting case]{\label{fig:wlnvpara39comp}\includegraphics[width=0.45\textwidth]{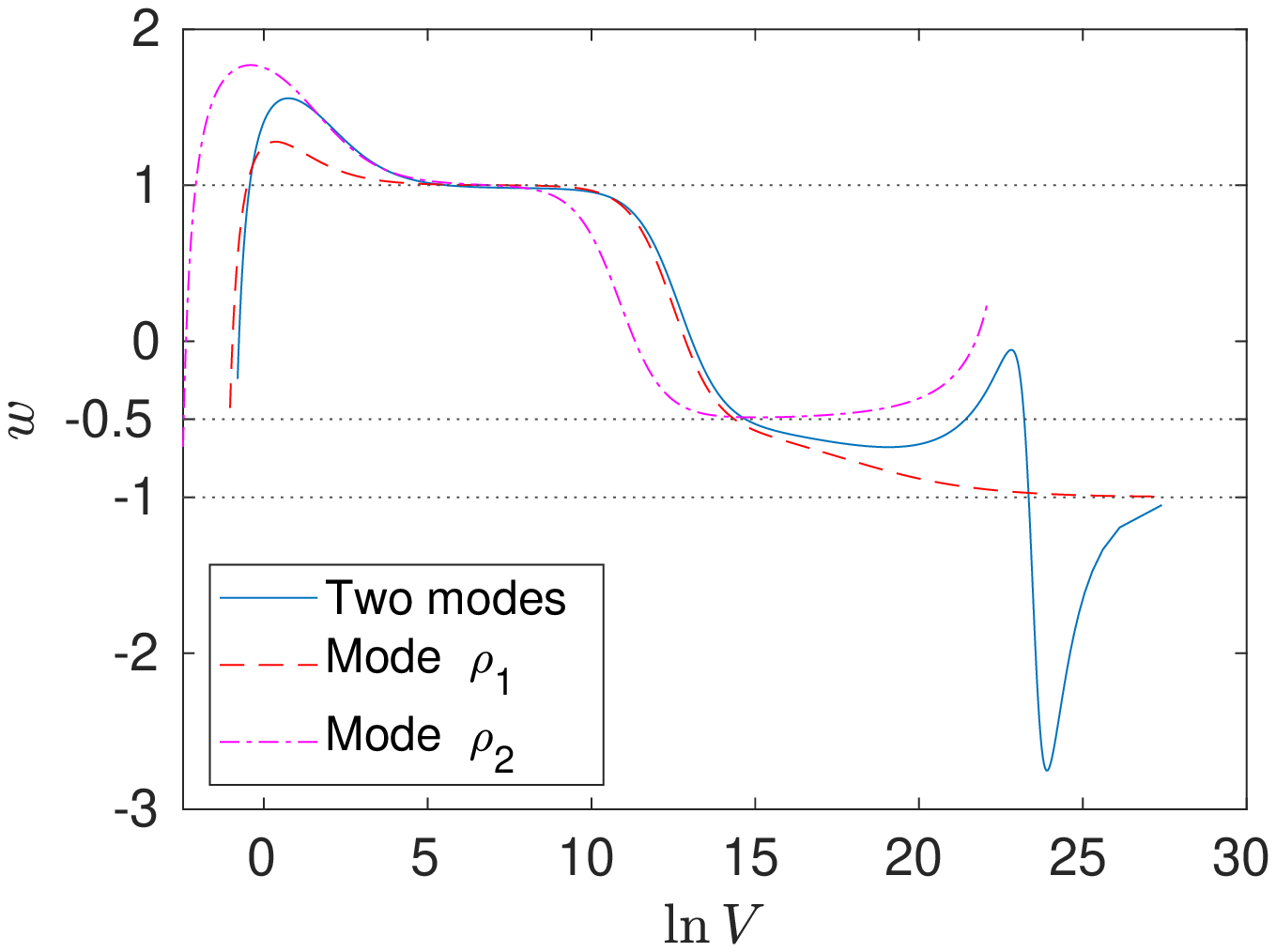}}
  \subfigure[~$w$ versus redshift in two modes case]{\label{fig:wzpara39}\includegraphics[width=0.45\textwidth]{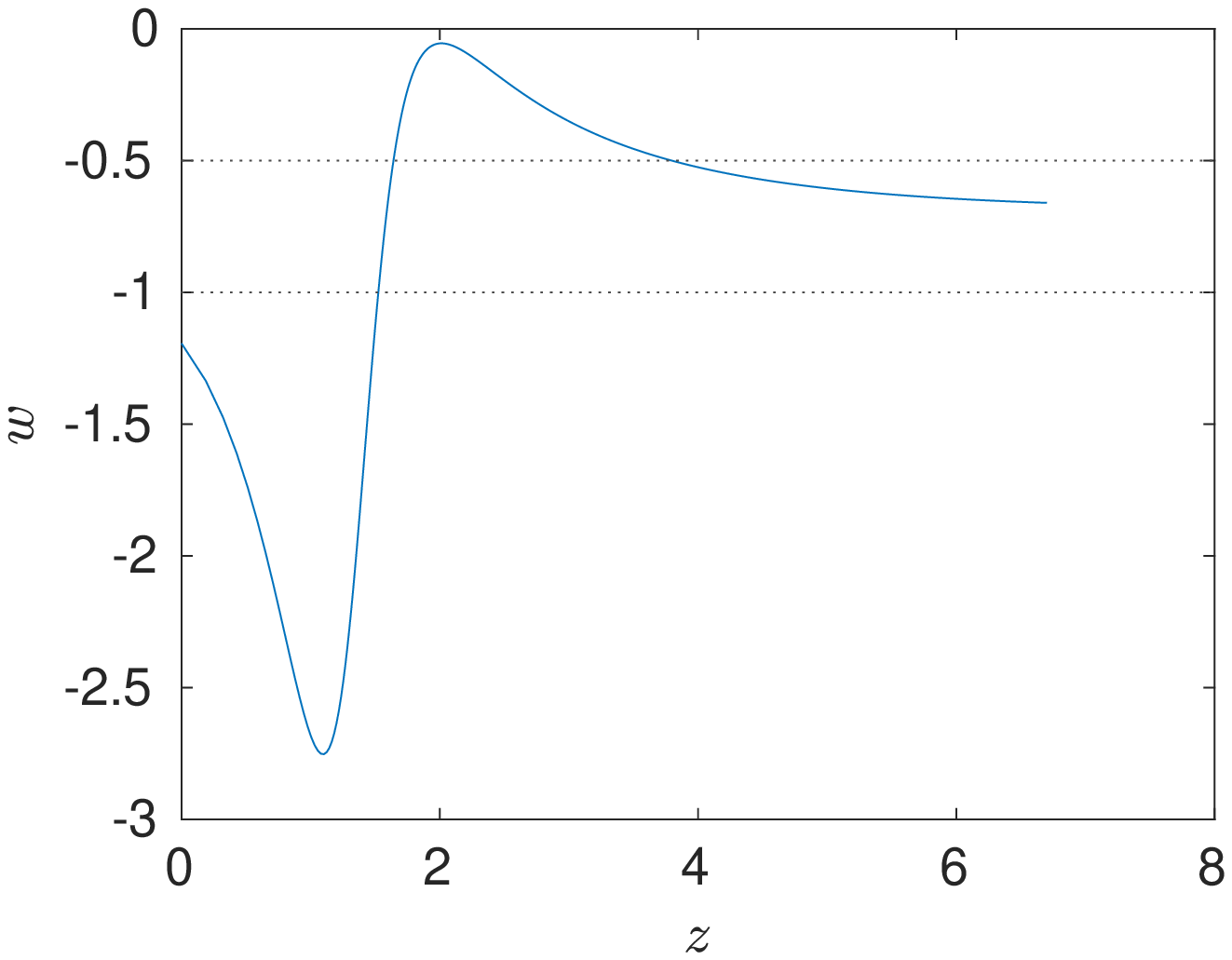}}
   \caption{The behaviour of $w$ in the interacting case. As in figure \ref{fig:wlnvfreepara31}, the blue sold line shows $w$ in two modes case, red dashed line shows single mode case with $\rho_1$, and magenta dash-dotted line shows single mode case with $\rho_2$. At large volume $w$ for $\rho_1$ and $\rho_2$ differs significantly as the couplings $\mu_1$ and $\mu_2$ have different signs. The two black dotted lines represent $w=-0.5$ and $w=-1$ respectively. In \ref{fig:wzpara39} we plot the behaviour of $w$ with respect to redshift $z$ in the two modes case. The redshift is defined by $z=a_0/a-1$, where scale factor $a=V^{1/3}$ and $a_0$ is its current value. Parameters are same as in figure \ref{fig:wlnvfreepara31} with additional ones are given by $\lambda_1=-10^{-8},~\mu_1=-\times10^{-12},~\lambda_2=-1.4757\times10^{-7},~\mu_2=1.2\times10^{-12}$ and $n_1=n_2=5,~n_1'=n_2'=6$.}
	  \label{fig:wintbehaviour}
\end{figure}

We compared the behaviour of $w$ in two modes case and single mode case in figure \ref{fig:wintbehaviour}. 


\section{Summary and outlook} \label{sec:summary}
In this work, we have analysed the emergent cosmological dynamics corresponding to the mean field hydrodynamics of quantum gravity condensates, within the (tensorial) group field theory formalism. 

In particular, we have extended previous analyses in the literature by studying the cosmological effects of fundamental interactions between TGFT quanta, the candidate \lq quantum constituents of spacetime\rq\ , and on the contributions from different quantum geometric modes associated to them. The general consequence of such interactions is to produce an accelerated expansion of the universe, which can happen both at early times, after the quantum bounce predicted by the model, and at late times. 

We have analysed in detail the properties of such acceleration, by recasting the dynamics of the universe volume in terms of an effective equation of state, encoding the details of the quantum gravity dynamics.

In the early universe right after the quantum bounce replacing the classical big bang singularity, the total volume is small and interaction terms remain subdominant, while we need to take into account all quantum geometric modes. In this regime, we studied whether the acceleration experienced by the universe right after the bounce could be long-lasting enough to have interesting cosmological consequences as a replacement for a later inflationary expansion. 
We were able to get an upper bound of the ratio $V_e/V_b$ between the volume at the end of acceleration phase and the beginning of the acceleration phase or the bounce. This bound is small under natural assumptions, i.e. the acceleration phase ends quickly after the bounce.

Away from the bounce, as long as the universe volume grows but interactions remain subdominant, one has a standard FLRW phase, whose precise duration depends on the value of the interaction coupling constants (in relation to the free part of the TGFT action). If instead interactions become relevant before such FLRW phase is reached, a long-lasting inflation-like expansion can be obtained, which however is not followed by a FLRW phase but by a collapsing phase (producing a cyclic universe).  

Further away from bounce, at larger values of the total volume, after the FLRW phase, the interaction terms for each mode become relevant and then dominate over free terms.

In the case of a single interaction term, we solve the equation explicitly. 

If the effective equation of state is mostly determined by a single mode, one has an effective equation of state $w=2-n/2$ for the interaction of order $n$. Then for $n\geq 5$, the expansion of universe is accelerating, approaching its asymptotic value from above. For $n\leq6$ this means that the phantom divide $w=-1$ cannot be crossed but only approached asymptotically, while for $n>6$ we have negative finite $w+1$, such that the reconstructed energy density of the fictitious field with such equation of state would diverge as volume grows, leading to a Big Rip singularity. 

If two modes determine the effective equation of state, on the other hand, the resulting cosmological evolution is much more interesting, and provides already, in fact, an observationally viable scenario. 
The effective $w$ approaches its asymptotic value from below, and the phantom divide $w=-1$ can be crossed without the need of introducing interactions of order higher than $6$. For $n=6$, we get $w<-1$ at large volume, and hence the energy density of the fictitious field will increase as volume grows, just as we expect for a phantom field. On the other hand, $w+1$ becomes infinitely small as $V\to\infty$, consequently the energy density of the fictitious field approaches to a finite value. Therefore, the Big Rip will not occur, rather, the universe will approach to a de Sitter spacetime asymptotically. 

The inclusion of more interaction terms for each mode further complicates the detailed late time expansion, allowing for example for several accelerated phases of different type, but does not change this asymptotic phantom-like behaviour.

Therefore, our main result is that, the emergent cosmological dynamics for TGFT condensates produces naturally a phantom-like dark energy dynamics at late times, compatible with cosmological observations and free of future singularities, purely out of quantum gravity effects without the need of any additional phantom matter.


Before turning to a broader outlook from our work, let us point out several aspects in which our analysis can be improved and its results sharpened. 

The approximation methods we used to solve the interacting equations could certainly be improved, in particular in the case in which several interaction terms are present. Such improvement could give more details about the interplay of different interaction terms for different modes and provide a better understanding of the potential range of cosmological dynamics of this class of models. Most important, we need to develop numerical as well as analytical techniques to be able to take into account the multitude of quantum geometric modes entering the TGFT quantum dynamics. While it is obviously true that we have barely scratched the potential richness of their emergent cosmological dynamics, we should also note that, even with two modes, the asymptotic value of $w$ is still the same as in the single mode case, only the way that $w$ approaches this value changed. Therefore we would expect that adding more modes would not change the fact that $w$ approaches the asymptotic value from below, crossing the phantom divide and thus still producing a phantom-like dark energy dynamics for large volume. In this sense, it is the step we have taken in this work, i.e. from one to two modes, that encodes the main qualitative features of such models for what concerns the late time evolution of our universe, and our results can be expected to be rather general and solid.

Another technical point where more work is needed concerns the form we have used for the TGFT interactions. As we noted, we have taken a rather phenomenological approach, by not working with any specific TGFT model but with a rather general expression, incorporating some aspects of known models in the isotropic restriction (for example, the fact that different spin modes decouple, as in the EPRL model), but not their detailed expression. This has the advantage of ensuring a certain degree of generality for our results. It should be complemented by a careful analysis of specific TGFT models (including the study of their renormalization group flow), to make sure that our expression captures their relevant features at this cosmological level, or to extract new ingredients that need to be added to the phenomenological expression, as potentially changing the resulting cosmological evolution.

As a basis for such effective phenomenological approach, we also used the mean-field approximation, which may not be trusted at late times, where the interactions become large (indeed, recent analyses confirm this worry \cite{Gielen:2019kae}). But, we emphasize again, in our work the only truly relevant ingredients are encoded in the choice of effective action. We used the simplest (mean field) approximation to it for simplicity, and for a closer contact with previous work in the literature, but one can easily consider a more general setting. The main point of our results is that including more than one mode in such effective action can indeed change the evolution of the universe, especially at late times, where the single mode is expected to be dominating. 

More precisely, in order to obtain the expression \eqref{eq:totalVcondensate} for the total volume, we used the mean-field approximation based on field coherent states. However, for more general states, we do not expect that the template for the derivation of relational volume observable and its dynamics would be much different. Like in ordinary quantum field theory, the generic quantum effective action for TGFTs is also a function of the effective mean field corresponding to the expectation value of the field operator in the true vacuum/ground state of the theory (rather than the simple coherent state we used), and a similar approximation in which such mean field is suitably peaked with respect to the relational clock would lead to the desired expression for corresponding observables as well. Therefore, it probably makes more sense to see the effective action we used as a simplified form of the quantum effective action of some interesting TGFT model for quantum gravity, after including (some) quantum corrections, rather than taking it literally as the mean field dynamics of a specific model and hoping that it is not spoiled by quantum corrections, despite the possible strong interactions.

Stepping into a more fundamental issue, our analysis, as well as the interpretation of its results, relied on the relational strategy for the definition of observables in a quantum gravity context (see \cite{Tambornino:2011vg, Hoehn:2019owq} and references cited therein), and in particular for a diffeomorphism invariant notion of temporal evolution. Most recent work on TGFT cosmology has adopted the same strategy. For example, both the expression for the effective equation of state and its physical interpretation at different values depends on the interpretation of the scalar degree of freedom we used as a clock as a free massless scalar field. This is consistent with all we currently know about the coupling of such fields in a TGFT (and discrete gravity or spin foam) formalism. However, much remains to be understood in this domain, i.e. matter coupling in this quantum gravity context, the construction of material reference frames and the detailed comparison with the corresponding constructions in classical gravitational physics. A more solid understanding of this issue at the interface between TGFT quantum gravity and the foundations of spacetime/gravitational physics will provide an even more solid take on the cosmological results we have obtained.

From an even broader perspective, our universe is way too simple to be fully realistic. In our analysis we only considered isotropic and homogeneous universes spacetime, thus ignored the effects from anisotropies and inhomogeneities, even at a perturbative level, on the evolution of the universe. 
Interesting work in both these directions have been done, in the TGFT cosmology literature \cite{Gielen:2018xph, Gielen:2017eco,Gerhardt:2018byq, deCesare:2017ynn}. The same is true, in fact, for the effects of thermal fluctuations of the TGFT condensates on the emergent cosmological evolution \cite{Assanioussi:2020hwf}.
With the same aim for a more realistic global picture of the universe evolution, even staying at the homogeneous level, we need to improve our analysis to include additional matter content, starting from general interacting scalar fields \cite{Li2017} but including then also the typical fluid components used in standard cosmological scenarios. 

However, beyond their effects on global cosmological evolution, a proper description of cosmological inhomogeneities is what is needed to make solid contact with cosmological observations and truly embed physical cosmology within our quantum gravity framework. This remains our main goal.

\acknowledgements

We thank Jibril Ben Achour, Luca Marchetti, Andreas Pithis, Yili Wang and Ed Wilson-Ewing for many useful discussions. We also appreciate several insightful comments from Steffen Gielen. DO acknowledges financial support from the Deutsche Forschung Gemeinschaft (DFG). XP also appreciates the financial support from China Scholarship Council. 

\appendix

\section{The effective equation of state} \label{sec:effw}

We want to define the equation of state of the content in the universe using only geometrical quantities. From the FLRW equation in a universe filled with different matter contents (represented by $i$)
\ieas{
  H^2=\frac{1}{3}\sum_i\rho_i,~\dot{H}=-\frac{1}{2}\sum_i(\rho_i+p_i)=-\frac{1}{2}\sum_i(1+w_i)\rho_i,
}
where $H=\frac{\dot{a}}{a}$ is the Hubble parameter, $\dot{~}$ represents derivative respect to comoving time, and $w_i=p_i/\rho_i$ is the equation of state for matter species. We can define an effective equation of state as
\iea{
  1+w&=& -\frac{2\dot{H}}{3H^2}.
}
In the relational time $\phi$, we have
\ieas{
  H=\frac{\dot{a}}{a}=\frac{1}{3}\frac{V'}{V}\dot{\phi}.
}
Using the fact that $\pi_\phi=\dot{\phi}V$ is a conserved quantity, we have $0=\ddot{\phi}V+\dot{\phi}\dot{V}=\ddot{\phi}V+\dot{\phi}^2V'$, and $\ddot{\phi}$ can be solved as
\ieas{
  \ddot{\phi}=-\frac{V'}{V}\dot{\phi}^2.
}
Therefore we have \cite{deCesare:2016rsf}
\ieas{
  \dot{H}&=& \frac{1}{3}\frac{\dd}{\dd\phi}\frac{V'}{V}\dot{\phi}^2+\frac{1}{3}\frac{V'}{V}\ddot{\phi} = \frac{1}{3}\frac{V''}{V}\dot{\phi}^2-\frac{1}{3}\left(\frac{V'}{V}\right)^2-\frac{1}{3}\left(\frac{V'}{V}\right)^2\dot{\phi}^2 = \frac{1}{3}\dot{\phi}^2\left[\frac{V''}{V}-2\left(\frac{V'}{V}\right)^2\right].
}
And the equation of state can be rewritten as
\iea{
  w=-\frac{2}{3}\frac{\dot{H}}{H^2}-1=-2\left[\frac{VV''}{(V')^2}-2\right]-1=3-\frac{2VV''}{(V')^2} \qquad . \label{eq:effwdefappendix}
}
When the evolution of the equation of state is know, the evolution of volume of universe can be recovered. To do this, we first introduce the relational Hubble parameter\footnote{Using the relation between volume and scale factor $V=a^3$, we see that the relation between Hubble parameter $H$ and the relational one $G$ is $\displaystyle H=\frac{\dot{a}}{a}=\frac{a'}{a}\dot{\phi}=\frac{G}{3}\dot{\phi}$. }
\iea{
  G=\frac{V'}{V} \qquad. \label{eq:relHubG}
}
Then the effective equation of state \eqref{eq:effwdefappendix} can be written by
\iea{
  w=1-\frac{2G'}{G^2} \qquad .
}
Suppose that $w$ is known, the equation is an ordinary differential equation of $G$ and can be solved by
\iea{
  G=\left(\int_{\phi_0}^\phi\frac{w(\chi)-1}{2}\dd\chi+\frac{1}{G_0}\right)^{-1},
}
where $G_0=G(\phi_0)$ is the initial value of $G$. Then the definition \eqref{eq:relHubG} of $G$ becomes a differential equation of volume $V$, and can be solved as
\iea{
  \ln V=\int_{\phi_0}^\phi\dd\kappa \left(\int_{\phi_0}^\kappa\frac{w(\chi)-1}{2}\dd\chi+\frac{1}{G_0}\right)^{-1}+\ln V_0,
}
with $V_0=V(\phi_0)$. Hence the evolution of volume respect to relational time $\phi$ is recovered.

\section{The consequences of the convergence of total volume $V$} \label{sec:convergentV}
In this appendix we consider how the convergence of $V$ will constrain parameters in our model. When $\phi$ is large, we will have 
\ieas{
  \sqrt{E_j^2+m_j^2Q_j^2}\cosh(2m_j\phi)-E_j>\sqrt{E_j^2+m_j^2Q_j^2}.
}
Therefore, if $V$ is convergent, the series
\iea{
  \sum_j\frac{V_j}{m_j^2}\sqrt{E_j^2+m_j^2Q_j^2} \label{eq:freeserieseqm}
}
must also be convergent. Then if $m_j$ is unbounded in the sense that $m_j\to\infty$ for $j\to\infty$, $V$ would certainly be divergent cause in terms with sufficient large $j$, we will have $\cosh(2m_j\phi)\to\infty$ for non-zero $\phi$. Therefore, the convergence of $V$ also requires bounded $m_j$.

Conversely, if series \eqref{eq:freeserieseqm} is convergent and $\forall j,~m_j\leq m$ with a given $m$, then
\ieas{
  \cosh(2m_j\phi)\leq\cosh(2m\phi),
}
which leads to the convergent of series
\ieas{
  \displaystyle \sum_j\left[\frac{V_j}{m_j^2}\sqrt{E_j^2+m_j^2Q_j^2}\cosh(2m\phi)\right]=\cosh(2m\phi)\sum_j\frac{V_j}{m_j^2}\sqrt{E_j^2+m_j^2Q_j^2},
}
since its right hand side is convergent according to our assumption. Therefore, series
\iea{
  \displaystyle \sum_j\left[\frac{V_j}{m_j^2}\sqrt{E_j^2+m_j^2Q_j^2}\cosh(2m_j\phi)\right]
}
converges as well. Furthermore, since $E_j<\sqrt{E_j^2+m_j^2Q_j^2}$, we see that $\displaystyle \sum_j\frac{V_jE_j}{m_j^2}$ is also convergent.

In conclusion, if we require $\rho_j'=0$ at the bounce for all $j$, then $V$ is convergent if and only if $\displaystyle \sum_j\frac{V_j}{m_j}\sqrt{E_j^2+m_jQ_j^2}$ converges and $m_j$'s are bounded. Just as we referred in section \ref{sec:freecondensate}.

\section{Behaviour of $\phi_{j\infty}$ in $n_j=4$ case for small $\lambda_j$} \label{sec:phiinfn4}
Here we consider the large $\rho_j$ behaviour for $n_j=4$ case, where we have an exact solution. In fact, for $n_j=4$, the solution of equation of motion \eqref{eq:rhoprimelambdamu} with $\mu_j=0$ can be expressed using elliptic functions. With the convention that $F(\phi,m)=\int_0^\phi\frac{1}{\sqrt{1-m\sin^2(\theta)}}\dd\theta$, we have the solution for a given mode $j$ with $\lambda_j<0$ \cite{Gradshteyn:1996table}
\iea{
  \phi&=& \sqrt{\frac{2}{-\lambda(\omega_3-\omega_1)}}F\left(\sin^{-1}\left(\sqrt{\frac{\rho_j^2-\omega_3}{\rho_j^2-\omega_2}}\right),\frac{\omega_2-\omega_1}{\omega_3-\omega_1}\right), \label{eq:phisollambdan4}
}
where $\omega_3>\omega_2>\omega_1$ are three real roots of the polynomial
\iea{
  P(\chi)&=& \chi^3-\frac{m^2}{2\lambda}\chi^2-\frac{E}{\lambda}\chi+\frac{2Q^2}{\lambda} \qquad , \label{eq:chipolyn4}
}
and the solution valids for $\rho_j>\sqrt{\omega_3}$. Note that $|\lambda_j|$ should be small enough such that the three roots of the polynomial \eqref{eq:chipolyn4} are all real. Setting $\rho_j\to\infty$ in the solution \eqref{eq:phisollambdan4}, we get the exact asymptotic value $\phi_{j\infty}$ in $n_j=4$ case
\iea{
  \phi_{j\infty}=\sqrt{\frac{2}{-\lambda_j(\omega_3-\omega_1)}}K\left(\frac{\omega_2-\omega_1}{\omega_3-\omega_1}\right). \label{eq:phiinfn4}
}
Now we consider the behaviour of this $\phi_{j\infty}$ for small $|\lambda_j|$. To do this, we need first find the approximate roots for the polynomial \eqref{eq:chipolyn4}. At the first order of $\lambda_j$, these roots are
\ieas{
  \omega_1&=& \frac{2m_j^2}{2\lambda_j}+\frac{2E_j}{m_j^2}+\frac{2E_j^2}{m_j^6}\lambda_j ,\nonumber \\
  \omega_2&=& \frac{Q_j^2}{E_j-\sqrt{E_j^2+m_j^2Q_j^2}}+\frac{Q_j^4\left(1+\frac{E_j}{\sqrt{E_j^2+m_j^2Q_j^2}}\right)}{4m_j^2\left(E_j-\sqrt{E_j^2+m_j^2Q_j^2}\right)^2}\lambda_j , \nonumber \\
  \omega_3&=& \frac{Q_j^2}{E_j+\sqrt{E_j^2+m_j^2Q_j^2}}+\frac{Q_j^4\left(1-\frac{E_j}{\sqrt{E_j^2+m_j^2Q_j^2}}\right)}{4m_j^2(E_j+\sqrt{E_j^2+m_j^2Q_j^2})^2} , \lambda_j
}
Then, putting these approximation of roots into equation \eqref{eq:phiinfn4}, we can further expand $\phi_{j\infty}$ with respect to small $\lambda_j$ using the expansion $K(x)\to\ln\frac{4}{\sqrt{1-x}}$ for $x\to1$, and we will obtain the same result as given by the corrected value \eqref{eq:phiinfncor} of $\phi_{j\infty}$.

\bibliographystyle{apsrev4-1}
\bibliography{./bib_library/library}

\end{document}